\documentclass[pdflatex,sn-mathphys]{sn-jnl}

\usepackage{graphicx}%
\usepackage{multirow}%
\usepackage{amsmath,amssymb,amsfonts}%
\usepackage{amsthm}%
\usepackage[title]{appendix}%
\usepackage{xcolor}%
\usepackage{textcomp}%
\usepackage{manyfoot}%
\usepackage{booktabs}%
\usepackage{algorithm}%
\usepackage{algorithmicx}%
\usepackage{algpseudocode}%
\usepackage{listings}%
\usepackage{mathtools}
\usepackage{latexsym}
\usepackage{bibentry}
\usepackage{natbib}
\usepackage{subcaption}

\usepackage{algcompatible}

\theoremstyle{thmstyleone}%
\theoremstyle{thmstyletwo}%
\newtheorem{example}{Example}%

\newenvironment{examplecont}[1]
  {\begin{trivlist}\item[\hskip\labelsep\bfseries Example~#1 (continued).]}
  {\end{trivlist}}
\newtheorem*{example*}{Example}

\theoremstyle{thmstylethree}%

\begin{document}

\title[Article Title]{Robust Bayesian methods using amortized simulation-based inference}


\author[1]{\fnm{Yuyan} \sur{Wang}}

\author[2]{\fnm{Michael} \sur{Evans}}

\author*[3]{\fnm{David J.} \sur{Nott}}\email{standj@nus.edu.sg}

\affil[1]{\orgdiv{Department of Statistics and Data Science}, \orgname{National University of Singapore}, \orgaddress{\street{ 6 Science Drive 2}, 
\postcode{117546}, 
\country{Singapore}}}

\affil[2]{\orgdiv{Department of Statistical Sciences}, \orgname{University of Toronto}, \orgaddress{\street{Ontario}, \city{Toronto}, \postcode{ M5S 3G3}, 
\country{Canada}}}

\affil*[3]{\orgdiv{Department of Statistics and Data Science}, \orgname{National University of Singapore}, \orgaddress{\street{ 6 Science Drive 2}, 
\postcode{117546}, 
\country{Singapore}}}


\abstract{Bayesian simulation-based inference (SBI) methods are used in statistical models where simulation is feasible but the likelihood is intractable. Standard SBI methods can perform poorly in cases of model misspecification, and there has been much recent work on modified
SBI approaches which are robust to misspecified likelihoods.  However, less attention has been given to the issue of inappropriate prior specification, which is the focus of this work.  
In conventional Bayesian modelling
there will often be a wide range of prior distributions consistent
with limited prior knowledge expressed by an expert.  
Choosing a single prior can lead to an inappropriate choice, possibly 
conflicting with the likelihood information.  Robust Bayesian methods, where a class 
of priors is considered instead of a single one, 
can address this issue. For each density in the prior class, a 
posterior can be computed, and the range of the resulting inferences is informative 
about posterior sensitivity to the prior imprecision. We consider density ratio classes for 
the prior and implement robust Bayesian SBI using amortized neural methods
developed recently in the literature. We also discuss methods for checking for conflict 
between a density ratio class of priors and the likelihood, and sequential updating 
methods for examining conflict between different groups of summary statistics. 
The methods are illustrated for several simulated and real examples.}




\maketitle
\section{Introduction}\label{sec:Intro}

There are many interesting statistical models where the likelihood is intractable.  If simulation of synthetic data from the model is feasible, it may still be possible to perform
Bayesian inference.  The field of simulation-based inference (SBI) deals with such models.  This paper develops robust Bayesian methods for SBI building
on recent amortized neural SBI methods in the literature.  Robust Bayesian approaches consider
Bayesian updating for every prior in some class rather than a single prior.  The range of
the resulting posterior
inferences is informative about posterior sensitivity to the prior ambiguity.  Computation of robust Bayesian inferences is challenging, but
one tractable approach uses so-called density ratio classes.  Our work considers robust Bayesian methods implemented with amortized SBI methods for density ratio classes.  We also develop
methods for checking for conflict between a density ratio prior class and the likelihood, and
methods for checking for conflict between subsets of data summary statistics.  

There are many different methods for simulation-based inference.  
One well-established approach is approximate Bayesian computation 
(ABC) \citep{tavare+bgd97,sisson+fb18intro}, which in its simplest form 
repeatedly simulates from the joint Bayesian model
for parameters and data, and accumulates parameter samples for which synthetic 
data is close enough to the observed data for some distance and tolerance.  
To make computation easier, 
the distance used in ABC is 
usually defined from low-dimensional summary statistics, and
the choice of these summary statistics is a crucial issue (see, for example, \cite{hoffmann+j26} for a recent discussion).  Another common 
SBI method is synthetic likelihood
\citep{wood10,price+dln16,frazier+ndk23}.  This approach 
approximates the likelihood by assuming
that the distribution of data summary statistics is Gaussian, and estimates 
summary statistic means and covariances using model simulation.   
The problem of estimating the posterior density from simulated data can also
be treated as one of flexible conditional density estimation, with neural posterior
estimation (NPE) \citep{papamakarios+m16,lueckmann+gbonm17,greenberg+nm19,radev+mvak22} being the most popular example of this approach.  Related neural likelihood
estimation (NLE) methods approximate the likelihood instead of the prior
\citep{papamakarios19,lueckmann+bkm19}, and approximations of likelihood ratios 
can also be developed
using flexible classifiers
\citep{cranmer+pl15,hermans+bl20,thomas+dckg21}.  Some recent
approaches approximate the posterior and likelihood simultaneously \citep{wiqvist+fp21,gloeckler+dm22,radev-etal-23}.  
A recent discussion of theoretical aspects of both NPE and NLE methods
is given by \cite{frazier+kdw24}, but the existing theory covers only the
case of a correctly specified model.  

Much recent SBI research has focused on the effects of misspecification on 
standard SBI methods.  However, the existing work mostly considers situations in which the
likelihood is misspecified.  A brief discussion of different approaches is given in Appendix A, and \cite{kelly+wfngd25} give a comprehensive recent review.  
An anonymous referee has also 
suggested the idea of combining the approaches discussed
here with generalized Bayesian methods, which
is an interesting direction which we do not pursue here.  
Instead we concentrate on neural conditional density estimation approaches 
for SBI with a probabilistic model and the 
issue of avoiding inappropriate choices of the  prior, and for understanding
the sensitivity
of posterior inferences to ambiguity when it is difficult to specify a single prior.  
To the best of our knowledge, there is no existing literature on robust Bayesian 
methods in this sense 
for models with intractable likelihood.  However, there is some work on checking for prior-data conflict in the conventional Bayesian setting of SBI 
with a single prior \citep{chakraborty+ne23}, and other authors have recognized the distinction between prior-data conflicts and misspecification
of the likelihood \citep{scmitt+bkr24}.  There is existing work on prior-data
conflict and imprecise probability (e.g., \citealp[]{walter+c16}, among others) 
but to the best of our knowledge 
not in the setting of density ratio classes specifically or for intractable
likelihood.  Further discussion is given in Section 5.  Our work makes three main
contributions.  The first is to implement
robust Bayes methods based on density ratio prior classes using amortized
Bayesian inference methods for SBI.  The second is to give methods for checking
for conflict between the likelihood and a density ratio prior class.  
Related to these checks, our third
contribution considers sequential updating for density ratio classes
to check for conflicts between subsets of summary statistics.  

In the next section we give some necessary background on robust Bayesian methods
using density ratio classes.  Section 3 discusses amortized neural methods for
SBI.  Section 4 discusses how to use the methods in Section 3 to implement
robust Bayesian methods in models with intractable likelihood, and  
Section 5 discusses conflict checking for density ratio classes and checking
for conflicts between subsets of summary statistics.  Section 6 discusses some
real and simulated examples and Section 7 gives some concluding discussion.

\section{Robust Bayes methods using density ratio classes}\label{ambiguity}

When a single prior is specified in a Bayesian analysis, some of
its characteristics will be chosen in an arbitrary way, since in 
practice the prior information
doesn't determine the prior uniquely. 
Expert knowledge can also be flawed, resulting in conflict between
prior and likelihood information.  This 
can compromise sound Bayesian inference, as conflicting information
should not be combined thoughtlessly.  Prior-data conflict is a form of
model misspecification that is distinct from any problem with the 
specification of the likelihood;  see \cite{evans+m06} and \cite{nott+wee16} for further
discussion.  

One way to avoid specification of a single prior
when that is difficult
is to take a robust Bayesian approach
\citep{walley91,berger94}.  
In robust Bayes a single prior is replaced by a set of priors.  For each prior
in the chosen set, we can compute a posterior distribution.  An important
question is how the class of priors should be defined so that it is expressive
of elicitation uncertainty but computationally tractable.  
Here we will use the so-called density ratio class
(also called intervals of measures in the original paper discussing
this class by
\cite{derobertis+h81}).  Useful overviews of this approach
with comparisons to other prior classes are given in \cite{berger90}
and \cite{rinderknecht+cbskr14}, and elicitation is discussed
in \cite{rinderknecht+br11}.  Bayesian updating and marginalization of a
density ratio class leads to another density ratio class.  
\cite{wasserman92}
proved that closure under Bayesian updating and marginalization 
characterize the density ratio
class. \cite{rinderknecht+cbskr14} discuss predictive inference 
for both deterministic and stochastic models.  
The rest of this section defines density ratio classes, and gives a summary
of their properties.

\subsection{Definition of the density ratio class}

To make our discussion easier we define some notation.  
For a model with parameter $\theta\in \Theta$ for data $y$, let 
$0\leq l(\theta)\leq u(\theta)$ be two functions, which we call
lower and upper bound functions, and assume that 
$$\int l(\theta)\,d\theta>0\;\;\;\text{and}\;\;\; \int u(\theta)\,d\theta<\infty.$$
Writing $\pi(\theta)$ for a possibly unnormalized density (i.e. 
a density that does not integrate to one), we follow
\cite{rinderknecht+cbskr14} and write $\hat{\pi}(\theta)$ for
the normalized verison of $\pi(\theta)$ when this exists.  
The density ratio class
with lower bound $l(\theta)$ and upper bound $u(\theta)$ is
\begin{align}
  \psi_{l(\theta),u(\theta)} := \left\{\hat{\pi}(\theta)=\frac{\pi(\theta)}{\int \pi(\theta)\,d\theta};
  l(\theta)\leq \pi(\theta)\leq u(\theta)\right\}. \label{drclass}
 \end{align}  

The functions $l(\theta)$ and $u(\theta)$ are 
bounds on the shape of a density in $\psi_{l(\theta),u(\theta)}$.  If an unnormalized
density can fit between the bounds, its normalized version is in $\psi_{l(\theta),u(\theta)}$.  
In the case where $l(\theta)>0$ for all $\theta\in\Theta$, an equivalent definition of $\psi_{l(\theta),u(\theta)}$ is
\begin{align}
  \psi_{l(\theta),u(\theta)}  = \Bigg\{\hat{\pi}(\theta): \frac{l(\theta)}{u(\theta')}&\leq \frac{\pi(\theta)}{\pi(\theta')}\leq \frac{u(\theta)}{l(\theta')} 
  \mbox{ for all $\theta,\theta'\in \Theta$}\Bigg\}.  \label{drclass2}
\end{align}
In \eqref{drclass2} the left-most inequality is equivalent to the right-most one 
by inverting ratios.  However, 
including the redundancy makes the implications of the definition clearer.  
The definition \eqref{drclass2} explains the name ``density ratio class"
first used in \cite{berger90}.  The equivalence of \eqref{drclass} and \eqref{drclass2}  
is demonstrated in Appendix B.  

From \eqref{drclass}, it is immediate that
$\psi_{l(\theta),u(\theta)}=\psi_{kl(\theta),ku(\theta)}$,
for any constant $k>0$.  Hence we could take either the lower or upper bound function to be a normalized density.  If we take $u(\theta)=\hat{u}(\theta)$, and 
write
\begin{align}
 r & =\frac{\int u(\theta)\,d\theta}{\int l(\theta)\,d\theta},  \label{rdefn}
\end{align}
then 
\begin{align}
  \psi_{l(\theta),u(\theta)}=\psi_{r^{-1}\hat{l}(\theta),\hat{u}(\theta)}. \label{normalized}
\end{align}
In visualizing density ratio classes later, we will normalize the upper
bound function.   It is helpful to introduce a simple example using density ratio classes to give
intuition for the definition. 
\begin{example}\label{ex:running} 
Suppose $y_1, \ldots, y_n$ are conditionally independent given $\theta \in \mathbb{R}^p$, with $y_i \mid \theta \sim N(\theta, \Sigma)$, where $\Sigma = \mathrm{diag}(\sigma_1^2, \ldots, \sigma_p^2)$ is known. The sample mean $\bar{y}$ is sufficient for $\theta$, with $\bar{y} \mid \theta \sim N(\theta, \Sigma/n)$. We write $\phi(x; \mu, \sigma^2)$ for the univariate normal density with mean $\mu$ and variance $\sigma^2$. We consider a density ratio class $\psi_{l(\theta),u(\theta)}$ for the prior on $\theta$ with bound functions
\[
l(\theta) = r^{-1}\prod_{j=1}^p \phi(\theta_j; \mu_{l,j}, \tau_{l,j}^2), \qquad u(\theta) = \prod_{j=1}^p \phi(\theta_j; \mu_{u,j}, \tau_{u,j}^2),
\]
where the upper bound is normalized to integrate to one
and has independent normal terms with means $\mu_{u,j}$ and variances
$\tau_{u,j}^2$, the lower bound has independent normal terms with means $\mu_{l,j}$ and variances $\tau_{l,j}^2$, and $r$ is the ratio \eqref{rdefn}. 
To have a valid density ratio class we require $l(\theta)\leq u(\theta)$ for all $\theta$. Let $J = \{j : \tau_{l,j}^2 < \tau_{u,j}^2\}$. Then $l(\theta)\leq u(\theta)$ for all $\theta$ if and only if for each $j \notin J$, $\tau_{l,j}^2 = \tau_{u,j}^2$ and $\mu_{l,j} = \mu_{u,j}$ and
\begin{equation}
\sum_{j \in J} \frac{(\mu_{u,j} - \mu_{l,j})^2}{\tau_{u,j}^2 - \tau_{l,j}^2} \;\leq\; 2\log r - \sum_{j \in J} \log\frac{\tau_{u,j}^2}{\tau_{l,j}^2}. \label{eq:final}
\end{equation}
The derivation of this condition is given in Appendix B. 
The requirement that the variances
are not smaller for the upper bound than the lower bound is quite intuitive, since if that
were not the case the bounds would cross eventually no matter how large $r$ is, 
violating the requirement $l(\theta)\leq u(\theta)$ for all $\theta$.  Prior
density ratio classes can accommodate priors with different tail behaviour.  When this occurs, the bound with heavier tail must be the upper bound.  \qedsymbol
\end{example}

In Example \ref{ex:running}, we have stated lower and upper bounds $l(\theta)$ and $u(\theta)$
which are valid, but 
how do we choose the bounds in practice?
Just as in a conventional Bayesian analysis where the prior needs to be 
elicited, for a robust Bayesian analysis using density ratio classes 
we must elicit $\hat{l}(\theta)$, $\hat{u}(\theta)$ and the constant $r$ in \eqref{rdefn}.  
For the case of univariate $\theta$, 
\cite{rinderknecht+br11} describe an elicitation procedure that begins by asking 
an expert for an interval of plausible values of quantiles for a set of fixed probabilities, and/or 
an interval of plausible cumulative distribution function values for a set of fixed quantiles.  
The elicitation of intervals captures the prior ambiguity 
of the expert.  
Next, some parametric families of densities are elicited to set the shape of the 
lower and upper bound functions.  The purpose of this is to eliminate unreasonable shapes 
for the densities in the class.  The parametrized bounds are then optimized to obtain a density 
ratio class that is as small as possible (i.e., with minimum $r$) while still respecting 
the elicited probability and quantile ranges.  Wider intervals reflecting more 
uncertainty in the expert's elicitation will result in larger $r$.  Multivariate elicitation 
is much more difficult than univariate elicitation, 
but \cite{rinderknecht+cbskr14} consider eliciting density ratio classes for 
each parameter separately.  They suggest a way to use these marginal prior
density ratio classes which is based on the closure property of 
such classes under Bayesian updating and maringalization, which is
discussed further below.  

\subsection{Lower and upper probabilities}

A density ratio class implies a range of probabilities for any event.
Suppose we are interested in the event $E\subseteq \Theta$, and 
for some density ratio class $\psi_{l(\theta),u(\theta)}$ we want 
the lower and upper probabilities $\underline{P}(E)$, $\overline{P}(E)$ 
for $E$, defined by 
\begin{align*}
(\underline{P}(E),\overline{P}(E)):=\Bigg(& \inf_{\hat{\pi}(\theta)\in \psi_{l(\theta),u(\theta)}} \int_E \hat{\pi}(\theta)\,d\theta,
\sup_{\hat{\pi}(\theta)\in \psi_{l(\theta),u(\theta)}} \int_E \hat{\pi}(\theta)\,d\theta\Bigg).
\end{align*}
It is easily shown (e.g. \citealt[Section 2.1]{rinderknecht+cbskr14}) that
\begin{align*}
\begin{aligned}
 (\underline{P}(E),\overline{P}(E))  = &\Biggl( 
 \frac{\int_E l(\theta)\,d\theta}{\int_E l(\theta)\,d\theta+\int_{E^c} u(\theta)\,d\theta},
 \frac{\int_E u(\theta)\,d\theta}{\int_E u(\theta)\,d\theta+\int_{E^c} l(\theta)\,d\theta}
\Biggl),
\end{aligned}
\end{align*}
where $E^c$ denotes the complement of $E$.  In terms of the normalized
densities $\hat{l}(\theta)$ and $\hat{u}(\theta)$ and with $r$ defined
in \eqref{rdefn}, we can write
\begin{align}
 (\underline{P}(E),\overline{P}(E))  =& \Biggl(
 \frac{\int_E \hat{l}(\theta)\,d\theta}{\int_E \hat{l}(\theta)\,d\theta+r \int_{E^c} \hat{u}(\theta)\,d\theta}, 
 \frac{\int_E \hat{u}(\theta)\,d\theta}{\int_E \hat{u}(\theta)\,d\theta+r^{-1}\int_{E^c} \hat{l}(\theta)\,d\theta}
 \Biggl).  \label{lower-and-upper}
\end{align}

\subsection{Closure under Bayesian updating and marginalization}

Two important properties of density ratio classes are closure under
Bayesian updating and marginalization.  \cite{wasserman92} 
demonstrated that the density
ratio class is the only class of densities possessing these properties.  
Invariance under Bayesian updating means the following.  If a
set of prior densities is a density ratio class, $\psi_{l(\theta),u(\theta)}$ say, 
and if we update each prior in $\psi_{l(\theta),u(\theta)}$ to its posterior density 
using a likelihood function $p(y|\theta)$, then the set of 
posterior densities is also a density ratio class.  We
write the resulting posterior density ratio class as
$\psi_{l(\theta|y),u(\theta|y)}$,
where $l(\theta|y)=l(\theta)p(y|\theta)$ and $u(\theta|y)=u(\theta)p(y|\theta)$.  
Later we will make use of the ratio of areas under the upper and lower
bound functions for the posterior density ratio class as a discrepancy
in a Bayesian predictive check, 
and it will be useful to have some notation for this.  
We write
\begin{align}
 r(y) & :=\frac{\int u(\theta|y)\,d\theta}{\int l(\theta|y) \,d\theta}=
 r \frac{\int \hat{u}(\theta)p(y|\theta)\,d\theta}{\int \hat{l}(\theta)p(y|\theta)\,d\theta},  \label{ry}
\end{align}
with the last equality following from \eqref{normalized}.  
The quantity $r(y)$ is a measure of how large the class of posterior
densities is.  Multiplying the lower
and upper bound functions by an arbitrary positive constant does not change
$r(y)$.    

Next we describe invariance of a density ratio class under marginalization.
Suppose we partition $\theta$ into two subvectors $\theta=(\theta_A^\top,\theta_B^\top)^\top$ and for $\hat{\pi}(\theta) \in \psi_{l(\theta),u(\theta)}$, write
$\hat{\pi}(\theta_A)$ for its $\theta_A$ marginal:
$$\hat{\pi}(\theta_A)=\int \hat{\pi}(\theta)\,d\theta_B.$$
The set of all densities $\hat{\pi}(\theta_A)$ for
$\hat{\pi}(\theta)\in \psi_{l(\theta),u(\theta)}$ is a density ratio class, $\psi_{l(\theta_A),u(\theta_A)}$, where
$$l(\theta_A)=\int l(\theta)\,d\theta_B,\;\;\;\;u(\theta_A)=\int u(\theta)\,d\theta_B.$$
For a posterior density ratio class $\psi_{l(\theta|y),u(\theta|y)}$, we will denote
the corresponding marginal posterior density ratio class for a subset $\theta_A$
of $\theta=(\theta_A^\top,\theta_B^\top)^\top$
as $\psi_{l(\theta_A|y),u(\theta_A|y)}$, where
$l(\theta_A|y)=\int l(\theta|y)\,d\theta_B,\;\;\;\;u(\theta_A|y)=\int u(\theta|y)\,d\theta_B.$

\begin{examplecont}{\ref{ex:running}}
Returning to Example \ref{ex:running}, write $l(\theta| \bar{y}) = l(\theta) p(\bar{y} \mid \theta)$ and $u(\theta| \bar{y}) = u(\theta) p(\bar{y} \mid \theta)$ for the lower and upper
bounds for the posterior density ratio class.  The normalized bounds are
\begin{equation}
\hat{l}(\theta \mid \bar{y}) = \prod_{j=1}^p \phi(\theta_j; \mu_{l,j}^\star, \tau_{l,j}^{\star 2}), \qquad \hat{u}(\theta \mid \bar{y}) = \prod_{j=1}^p \phi(\theta_j; \mu_{u,j}^\star, \tau_{u,j}^{\star 2}), \label{eq:posterior-bounds}
\end{equation}
with standard Gaussian updates coordinate-wise 
\[
\tau_{l,j}^{\star 2} = (\tau_{l,j}^{-2} + n\sigma_j^{-2})^{-1}, \qquad \mu_{l,j}^\star = \tau_{l,j}^{\star 2}\bigl(\tau_{l,j}^{-2}\mu_{l,j} + n\sigma_j^{-2}\bar{y}_j\bigr),
\]
and similarly for $(\mu_{u,j}^\star, \tau_{u,j}^{\star 2})$. The ratio \eqref{ry} is
\begin{equation}
r(\bar{y}) = r \cdot \prod_{j=1}^p \frac{\phi\bigl(\bar{y}_j; \mu_{u,j}, \tau_{u,j}^2 + \sigma_j^2/n\bigr)}{\phi\bigl(\bar{y}_j; \mu_{l,j}, \tau_{l,j}^2 + \sigma_j^2/n\bigr)}. \label{eq:r-star}
\end{equation}

For marginal posterior density ratio classes, partition $\theta = (\theta_A^\top, \theta_B^\top)^\top$ with $\theta_A = (\theta_j)_{j \in A}$ for some $A \subseteq \{1, \ldots, p\}$. The marginal posterior class for $\theta_A$ is the density ratio class $\psi_{l(\theta_A|\bar{y}), u(\theta_A|\bar{y})}$ with normalized bounds
\begin{equation}
\hat{l}(\theta_A \mid \bar{y}) = \prod_{j \in A}\phi(\theta_j; \mu_{l,j}^\star, \tau_{l,j}^{\star 2}), \qquad \hat{u}(\theta_A \mid \bar{y}) = \prod_{j \in A}\phi(\theta_j; \mu_{u,j}^\star, \tau_{u,j}^{\star 2}), \label{eq:marginal-bounds}
\end{equation}
so we can take $l(\theta_A|\bar{y})=r(\bar{y})^{-1}\hat{l}(\theta_A\mid \bar{y})$, 
$u(\theta_A|\bar{y})=\hat{u}(\theta_A|\bar{y})$. Derivations of the above
results are given in Appendix B. \qedsymbol
\end{examplecont}

Now that we have explained the closure of density ratio classes under
Bayesian updating and marginalization, we return to the issue of
elicitation.  \cite{rinderknecht+cbskr14} consider multivariate parameter
problems, and elicit prior density ratio classes for each parameter separately
using the approaches in \cite{rinderknecht+br11}.  
Suppose we have a density ratio class for one of the parameters, 
say $\theta_1$.  We can combine this density ratio class for $\theta_1$ with fixed independent 
priors for the other parameter components, to examine the effects of ambiguity
in the prior for $\theta_1$.  
\citet{rinderknecht+cbskr14} demonstrate how to construct a joint prior density ratio class in this situation 
such that the corresponding 
marginal posterior 
density ratio class for $\theta_1$ is the same as in the univariate analysis where the 
components $\theta_j$, $j\neq 1$, are integrated out from the likelihood using the fixed 
priors for those components.  
See \cite{rinderknecht+cbskr14}[Section 2.3, equation (29)].  One can also specify density 
ratio classes for all univariate parameters and combine the lower and upper bound 
values independently to obtain a joint density ratio class, but this results in 
conservative uncertainty quantification if the number of parameters is large. 
In this work, we focus mostly on the computational aspects of density ratio 
classes with SBI, but improving elicitation methods is a fruitful direction for 
future work.  The conflict checking methods of Section 5 can be helpful 
in detecting problems with the elicitation of a prior density ratio class.
For an analyst without expert knowledge, if a precise prior has been
specified, it is reasonable to use density ratio classes for sensitivity
analysis, where we specify bounds in the prior density ratio class that
represent a set of reasonable perturbations of a precise prior, and then
examine how the bounds change in the posterior density ratio class.  We can
visualize the effects of ambiguity on the parameters by looking at the 
marginal bounds for the posterior density ratio classes for each parameter. 

In example 1, we were able to obtain lower and upper bounds for joint and marginal
density ratio classes explicitly.  Later we approximate posterior densities
using conditional normalizing flows, where there is an explicit density
for the joint posterior density approximation, but not for marginal densities.  
This means that in summarizing marginal posterior density ratio class bounds, 
we need to resort to simulation from the approximation (which is easily done
directly) and kernel density estimation.

{\bf Notation:}  It is helpful to explain the logic behind notational
conventions used throughout the rest of the paper.  We use the usual conditioning
notation $|$ for conditional densities, but for lower bounds and upper bounds
the conditional densities may be unnormalized ones.  Sometimes
we wish to indicate dependence of a density or conditional density on
information which is not in the form of the value of a random variable.  Here
the information is put after a semi-colon.  For example, later we consider
predictive densities where we want to show the dependence on the prior.  
A predictive density obtained by integrating out parameter $\theta$ using a density
$\hat{\pi}(\theta)$ in $p(y|\theta)$ would be denoted $p(y;\hat{\pi})$.  
Subscripts in density ratio classes are the lower and upper bounds defining them.

\subsection{Prediction}

\cite{rinderknecht+cbskr14} considered how a density ratio class defined on
parameters propagates in Bayesian predictive inference for 
data $y'$.  
\cite{rinderknecht+cbskr14} consider both deterministic and stochastic
prediction.  For the case where predictions are deterministic given 
$\theta$, the predictive
densities propagated from a density ratio class on the parameter
space are a density ratio class on predictive space.  This follows
from the closure under marginalization property discussed above.

Next, consider stochastic prediction, where
the data to be predicted has density $p(y' |\theta)$ given $\theta$.  
Suppose we have a density ratio class $\psi_{l(\theta),u(\theta)}$ of densities defined 
on the parameter space $\Theta$.  For $\hat{\pi}(\theta)\in \psi_{l(\theta),u(\theta)}$, 
consider the prior predictive density for $y'$ defined as
\begin{align}
  p(y';\hat{\pi}) & := \int p(y'|\theta) \hat{\pi}(\theta)\;d\theta.  \label{predictive}
\end{align}
Later we will also consider extensions to settings where there is previously 
observed data $y$ say and $y'$ is not conditionally independent of it given $\theta$, 
and we define
\begin{align*}
p(y'|y;\hat{\pi}) & :=\frac{p(y',y;\hat{\pi})}{p(y;\hat{\pi})}.
\end{align*}
The set of $p(y';\hat{\pi})$ for all $\hat{\pi}(\theta)\in \psi_{l(\theta),u(\theta)}$ is
not a density ratio class, but it is contained in the density ratio class
\begin{align}
 \psi_{l(y'),u(y')}, \label{predictivedrclass}
\end{align}
where
$$l(y')=\int l(\theta)p(y'|\theta)\,d\theta \;\;\;\;
u(y')=\int u(\theta)p(y'|\theta)\,d\theta.$$
There can be predictive densities in \eqref{predictivedrclass} that may not be obtained
from \eqref{predictive} for some $\hat{\pi}\in \psi_{l(\theta),u(\theta)}$;  \citet[Section 2.4.1]{rinderknecht+cbskr14} give an example.
We can use the density ratio class \eqref{predictivedrclass} to give
conservative upper and lower predictive probabilities.  
It is immediate that
$$\psi_{l(y'),u(y')}=\psi_{(r)^{-1} \hat{l}(y'),\hat{u}(y')}.$$
Later we will consider density ratio classes of prior predictive densities based
on the observed $y$, as given by $\psi_{l(y),u(y)}$.  We will also consider prior predictive
densities of data summary statistics, $S=S(y)$, and in this case we
write $\psi_{l(S),u(S)}$.

\section{Amortized inference for SBI}

Next we discuss amortized inference for SBI in the conventional 
Bayesian framework.  The methods we describe here are used in the next
section to perform robust Bayesian SBI computations for density ratio classes.

Suppose we have data $y_{\text{obs}}$ and a model for it 
with density $p(y|\theta)$ where $\theta\in \Theta$ is an unknown parameter.  
The likelihood $p(y_{\text{obs}}|\theta)$ is intractable, and we consider neural
SBI methods for inference about $\theta$.   There are many neural SBI methods 
in the literature as discussed in the introduction;  
here we focus on amortized methods, 
which after training are able to produce
posterior approximations for arbitrary data (not just $y_{\text{obs}}$) at
minimal additional computational cost.  In our work, we have used
the JANA package \citep{radev-etal-23}, which builds on the
BayesFlow approach of \cite{radev+mvak22}, with the latter performing
only posterior estimation, while the former approximates both posterior
and likelihood.  Most neural SBI methods involve the sequential learning
of a proposal distribution over many rounds to focus the simulation effort on
parts of the parameter space likely to produce synthetic data similar to
the observed data.  Although these methods are not amortized, they can
often be thought of as amortized methods if a single round of training is performed
with a proposal given by the prior.  Amortized SBI 
is a fast-moving field \citep{gloeckler+dwwm24,zammit+sh24,chang+lhrka25}  
and our work makes no new contribution to it.  Our focus instead 
is on using amortized methods
to implement robust Bayesian inference with density ratio classes, and for calibrating
checks for prior-data conflict.  The latter task requires computing posterior quantities
many times for different simulated data, which amortized methods can do efficiently.

The methods described in \cite{radev-etal-23} are well-suited 
for robust Bayesian computations with density ratio classes, 
since they provide methods for approximating both marginal likelihoods
and posterior densities, and this can be exploited for 
computing quantities such as $r(y)$ defined
at \eqref{ry}.  \cite{rinderknecht+cbskr14} describe a variety of
approaches to computation with density ratio classes, some of which do not
require marginal likelihood evaluations, or only require estimation
of the posterior density for a single prior density, but we found these alternatives
to be more difficult numerically than those described in the next section.  
Another strength of the methodology in \cite{radev-etal-23} is the development of
simulation-based calibration (SBC)
\citep{talts+bsvg18} methods adapted to the setting of amortized inference 
where both posterior and likelihood approximations are learnt (so-called 
joint simulation-based calibration, JSBC).   

We briefly describe here the JANA methodology in the simplified case where
summary statistics are not learnt from the data.  
Learnt summary 
statistics are useful in many settings, but can be hard to interpret, and later
our summary statistic checks are more insightful when the summaries are specified
by the user so that they are interpretable. 
For data $y\in \mathcal{Y}$, we write $S:\mathcal{Y}\rightarrow \mathcal{S}$ for a summary
statistic mapping.  In our examples, $\mathcal{Y}\subseteq \mathbb{R}^n$ and $\mathcal{S}\subseteq \mathbb{R}^d$, where $n$ is the sample size and $d$ 
is the summary statistic dimension, and generally $d\ll n$.  The observed summary
statistic value is written $S_{\text{obs}}=S(y_{\text{obs}})$.  Parametrized
approximations are used for both the posterior and likelihood.  
For some parameter $\varphi$, we write
$q_\varphi(\theta|S)$ for a parametric approximation to the summary statistic posterior
$p(\theta|S)$ valid for all $S$, and for some parameter $\gamma$ we write
$q_\gamma(S|\theta)$ for a parametric approximation to the sampling density $p(S|\theta)$ of the summary statistic,
valid for all $\theta$.  How these families of 
posterior and likelihood approximations are parametrized
in a flexible way is discussed further below.  For a given parametrization, 
the final approximations for likelihood and posterior are
$q_{\gamma^*}(S|\theta)$, $q_{\varphi^*}(\theta|S)$, where 
\begin{align}
  (\varphi^*,\gamma^*) 
   = \arg \min_{\varphi,\gamma} E_{p(\theta)p(S|\theta)}\bigg\{-\log q_\varphi(\theta|S) 
   -\log q_{\gamma}(S|\theta)\bigg\}.   \label{objective}
\end{align}

In practice, the expectation needs to be approximated by an 
average over simulated samples, and a 
a penalty can also be added to encourage the prior predictive
density of summaries to be close to standard normal when summary statistics are learnt.  This can be useful in model checking \citep{scmitt+bkr24} for the detection
of unusual summary statistic values with respect to the prior.  Optimizing a 
simulation approximation
of \eqref{objective} without any penalty learns a 
parametric approximations to the posterior
and likelihood simultaneously using maximum likelihood for the simulated 
data.  

We choose
$q_\varphi(\theta|S)$ and $q_{\gamma}(S|\theta)$ to be
conditional normalizing flows (see, for example, \cite{papamakarios19} for a review of normalizing flows).  For an $\mathbb{R}^p$-valued random variable
$\theta$ having a distribution depending on $S$, suppose that
$T_{\varphi,S}:\mathbb{R}^p\rightarrow \mathbb{R}^p$ is a smooth invertible
mapping and that $T_{\varphi,S}(\theta)\sim N(0,I_p)$ for every $S$, where
$I_p$ denotes the $p\times p$ identity matrix.  By the
invertibility of $T_{\varphi,S}$, and using a change of variables, the implied
density of $\theta$ conditional on $S$ is
$$q_{\varphi}(\theta|S)=\phi_p(T_{\varphi,S}(\theta))\left| \det \frac{\partial T_{\varphi,S}}{\partial \varphi}\right|,$$
where $\phi_p(\cdot)$ denotes the $p$-dimensional standard normal density.  

This gives a parametrized form for the posterior density given $S$, for a suitable
family of transformations $T_{\varphi,S}$ with the tuning parameter $\varphi$.  
Similarly, for an $\mathbb{R}^d$-valued random variable $S$ with a distribution
depending on $\theta$, suppose that $H_{\gamma,\theta}:\mathbb{R}^d\rightarrow
\mathbb{R}^d$ is a smooth invertible mapping such that $H_{\gamma,\theta}(S)\sim N(0,I_d)$ for every $\theta$.  By the invertibility of $H_{\gamma,\theta}$ and using
a change of variables, the density of $S$ conditional on $\theta$ is
$$q_{\gamma}(S|\theta)=\phi_d(H_{\gamma,\theta}(S))
\left| \det \frac{\partial H_{\gamma,\theta}}{\partial \gamma}\right|.$$
This gives a parametrized form for the sampling density of $S$ given $\theta$.
The invertible mappings $T_{\varphi,S}$ and $H_{\gamma,\theta}$ are constructed
as compositions of simpler transformations (``flows''), 
for which there are many standard choices, 
such as real NCP \citep{dinh+db16} and neural spline flows \citep{durkan+bmp19}, among many
others.  Often the parameters in the flows are weights for neural networks.  In the flow construction it is important to ensure invertibility as well as easy computation of Jacobian determinants in the expressions
for $q_{\varphi}(\theta|S)$ and $q_{\gamma}(S|\theta)$.  
It is straightforward to obtain independent draws from a distribution defined 
through a normalizing flow from their definition in terms of a transformation
of a simple density.

\section{Implementing robust Bayes for SBI}

A common method of summarizing the posterior density in a conventional
Bayesian analysis is to plot the marginal densities.  Similarly, in robust Bayes
with density ratio classes, we can plot
the upper and lower bound functions of the density ratio classes for the marginal
posteriors.   
Let $\psi_{l(\theta),u(\theta)}$ be a prior density ratio class.  
Write $\hat{\pi}(\theta|y)$ for the posterior density obtained by updating
$\hat{\pi}(\theta)$ using likelihood $p(y|\theta)$ and Bayes' rule, and
write $\hat{\pi}(\theta_j|y)$ for its $\theta_j$ marginal density.
By the invariance under
Bayesian updating and marginalization properties, the set of densities
$$\{\hat{\pi}(\theta_j|y): \hat{\pi}(\theta)\in \psi_{l(\theta),u(\theta)}\},$$
is a density ratio class, with lower and upper bound functions

\begin{align}
 l(\theta_j;y)= & r(y)^{-1}\hat{l}(\theta_j|y)<\hat{u}(\theta_j|y)=u(\theta_j;y), \label{lowupthetaj}
\end{align}
\noindent
where $r(y)$ is defined in equation \eqref{ry}, 
$\hat{l}(\theta_j|y)$ and $\hat{u}(\theta_j|y)$ are $\theta_j$ marginal
posterior densities for priors $\hat{l}(\theta)$ and 
$\hat{u}(\theta)$ respectively,
and we have chosen to define the upper bound as a normalized
density function.

To compute the lower and upper bound functions in \eqref{lowupthetaj}, 
we need $\hat{l}(\theta_j|y)$, $\hat{u}(\theta_j|y)$ and
$r(y)$.  In the likelihood-free setting with
summary statistics $S$, we approximate $\hat{l}(\theta_j|y)$ by
first simulating data $Z_i^l=(\theta_i^l,S_i^l)$, $i=1,\dots, n$, where
the $Z_i^l$ are drawn independently from density $\hat{l}(\theta)p(S|\theta)$.  
Using the methods discussed in Section 3, we can obtain from this
training data approximations
$\widetilde{l}(\theta|S)$ and $\widetilde{l}(S|\theta)$ for the posterior
density of $\theta$ given $S$ and sampling density of $S$ given $\theta$
respectively when the prior is $\hat{l}(\theta)$.  
For the prior $\hat{u}(\theta)$ derived from the upper bound, 
we similarly obtain approximations $\widetilde{u}(\theta|S)$ and
$\widetilde{u}(S|\theta)$ for the posterior
density of $\theta$ given $S$ and sampling density of $S$ given $\theta$
respectively when the prior is $\hat{u}(\theta)$.  So by simulation
of samples from $\widetilde{l}(\theta|S)$ and $\widetilde{u}(\theta|S)$ we
can construct approximations $\widetilde{l}(\theta_j|S)$ 
and $\widetilde{u}(\theta_j|S)$ to the marginal posterior densities $\hat{l}(\theta_j|y)$
and $\hat{u}(\theta_j|y)$ by using kernel density estimates for the $\theta_j$ samples.

We also need to approximate the ratio 
\begin{align}
  r(y) & = \frac{\int u(\theta|y)\,d\theta}{\int l(\theta|y)\,d\theta} =r\frac{\int \hat{u}(\theta)p(y|\theta)\,d\theta}{\int \hat{l}(\theta)p(y|\theta)\,d\theta} = r\frac{p(y;\hat{u})}{p(y;\hat{l})}, \label{rycomputable}
\end{align}
where $r$ is defined in \eqref{rdefn} and is known from the elicitation
of $\psi_{l(\theta),u(\theta)}$.
Recalling the definition in equation \eqref{predictive},  
$p(y;\hat{l})$ and $p(y;\hat{u})$ are the values of the prior predictive densities at $y$ for priors 
$\hat{l}(\theta)$ and $\hat{u}(\theta)$ respectively.  
In the likelihood-free case, where we use summary statistics $S$, 
we will extend our previous notation and write  
\begin{align}
 r(S) & = r\frac{p(S;\hat{u})}{p(S;\hat{l})},  \label{rScomputable}
\end{align}
where a prior predictive density for $S$, based on 
distribution $\hat{\pi}(\theta)$ on the parameter space, is written as
$p(S;\hat{\pi})$.  

To approximate $p(S;\hat{l})$ and $p(S;\hat{u})$ in \eqref{rScomputable}, 
we know from Bayes' rule that
$$\hat{l}(\theta|S) = \frac{\hat{l}(\theta)p(S|\theta)}{p(S;\hat{l})} \;\;\;\;\text{and}\;\;\;\;
\hat{u}(\theta|S) = \frac{\hat{u}(\theta)p(S|\theta)}{p(S;\hat{u})}.$$
Rearranging these expressions, 
\begin{align}
 & p(S;\hat{l}) = \frac{\hat{l}(\theta)p(S|\theta)}{\hat{l}(\theta|S)}
\;\;\;\;\text{and}\;\;\;\;
 p(S;\hat{u}) = \frac{\hat{u}(\theta)p(S|\theta)}{\hat{u}(\theta|S)}. \label{lowerml}
\end{align}
These expressions hold for any choice of $\theta$.  Using the same
$\theta$ in both expressions at \eqref{lowerml} to compute \eqref{rScomputable},
\begin{align}
  r(S) & = r\frac{\hat{u}(\theta)\hat{l}(\theta|S)}{\hat{l}(\theta)\hat{u}(\theta|S)},
  \label{diffml}
\end{align}
where the likelihood term cancels out when taking the ratio. 
Replacing $\hat{l}(\theta|S)$ and $\hat{u}(\theta|S)$ by normalizing
flow approximations $\widetilde{l}(\theta|S)$ and $\widetilde{u}(\theta|S)$ we obtain
the approximation
\begin{align}
  \widetilde{r}(S) & := r\frac{\hat{u}(\theta)\widetilde{l}(\theta|S)}{\hat{l}(\theta)\widetilde{u}(\theta|S)}.
  \label{diffmlapprox}
\end{align}
 We have experimented with choosing $\theta$ 
in computing \eqref{diffmlapprox} as the 
posterior mean  of $\widetilde{u}(\theta|S)$.  
However, in estimating log marginal likelihood values for amortized
Bayesian inference it is recommended \citep{kucharsky+mhrb25} to average over posterior samples.  In this
case, we can consider the approximation 
\begin{align} 
  \bar{r}(S) & := r \prod_{m=1}^M \left\{\frac{\hat{u}(\vartheta_m)\widetilde{l}(\vartheta_m|S)}{\hat{l}(\vartheta_m)\widetilde{u}(\vartheta_m|S)}\right\}^{1/M} \label{diffmlbar}
\end{align}
which is a geometric average of terms of the form \eqref{diffmlapprox} 
for samples $\vartheta_m$, $m=1,\dots, M$ from $\widetilde{l}(\theta|S)$. 
This geometric average corresponds to an arithmetic average on the log scale.

 Now that we have discussed how to implement robust Bayesian methods
with density ratio classes, it is worth comparing the computational demands
relative to a similar conventional Bayesian analysis.   
The robust analysis is similar to doing a conventional Bayesian analysis
twice, with priors $\hat{l}(\theta)$ and $\hat{u}(\theta)$.  If we are using
amortized methods and obtain explicit expressions for posterior density
approximations, estimating marginal likelihoods and posteriors for different
data (which is used later for conflict checking) is relatively cheap computationally.
 
\section{Prior-data conflict checking}\label{sec:checking}

Our next goal is to devise checks for whether a density ratio class of priors 
is in conflict with the likelihood.  
The checks we develop are extensions of the checks used in 
conventional Bayesian analysis with a single prior, and we explain these 
first.  

\subsection{Conventional Bayesian predictive checks}

We use similar notation to previous sections, with 
$\theta$ a parameter in a statistical model for data $y$.  The density
of $y$ is $p(y|\theta)$, and we consider in this subsection 
conventional Bayesian inference with
a prior density $p(\theta)$ for $\theta$.  The observed value of $y$ is
$y_{\text{obs}}$.  The posterior density is 
$p(\theta|y_{\text{obs}})\propto p(\theta)p(y_{\text{obs}}|\theta)$.
Bayesian model checking is usually performed through
Bayesian predictive checks, which require the choice of a discrepancy
$D(y)$, and a reference density $m(y)$ for the data.  
The check examines whether the observed discrepancy $D(y_{\text{obs}})$ is
surprising or not under the assumed model.  To measure surprise, the
observed discrepancy is calibrated by computing a tail probability, 
sometimes referred
to as a Bayesian predictive $p$-value, 
\begin{align}
  p & = P(D(y)\geq D(y_{\text{obs}})),   \label{tailprob}
\end{align}
for $y\sim m(y)$, where it is assumed that $D(y)$ has been defined in
such a way that a larger value is considered more surprising.

The discrepancy and reference distribution used should depend on what
aspect of the model we wish to check.  When the goal is to check the likelihood
component of the model, it is popular to choose $m(y)$ as 
the posterior predictive density of a replicate observation \citep{guttman67,gelman+ms96}:
$$p(y|y_{\text{obs}})=\int p(\theta|y_{\text{obs}})p(y|\theta)\,d\theta.$$
Although easy to use, posterior predictive $p$-values 
are not necessarily uniform or otherwise having a known distribution when
the model is correct, and this lack of calibration means it can be hard
to identify when such a check has produced a surprising result.  
Lack of calibration is often symptomatic of an insufficiently thoughtful choice of $D(y)$, but alternatives to the posterior predictive $p$-value with better 
calibration properties have been explored \citep{bayarri+b00,moran+br23}.  

Most relevant to the present work is 
checking for prior-data conflict, where we
consider checking the prior $p(\theta)$ rather than the likelihood 
$p(y|\theta)$.  By
``checking the prior" here, we mean considering a Bayesian predictive check
which will alert us if the information in the prior and the 
likelihood are contradictory.  This happens if $p(\theta)$ puts all its mass in
the tails of the likelihood.  These conflicts are important to detect, because
they can result in prior sensitivity for inferences of interest \citep{allabadi+e17}, and may alert us to an inadequate understanding of the model and its parametrization.  For prior-data
conflict checking, an appropriate reference density $m(y)$ is the prior
predictive density \citep{box80}, 
$$p(y)=\int p(\theta)p(y|\theta),$$
since we wish to see if the observed likelihood is unusual compared
to the likelihood for data generated using the prior density for the parameters.  
\cite{box80} suggested prior predictive checking for criticism of both likelihood 
and prior, but \cite{evans+m06} argued that prior predictive checks based on a minimal
sufficient statistic are appropriate for checking for prior-data conflict, while alternative
methods are more appropriate for criticizing the likelihood.  
Various choices of $D(y)$ can be considered for summarizing the
likelihood in a prior predictive check, such as prior predictive density
values for 
exactly or approximately sufficient statistics, or prior-to-posterior 
divergences \citep{evans+m06,nott+wee16}.  Extensions
to hierarchical conflict checking methods are also discussed by
these and other authors (e.g. \citealt{marshall+s07,bayarri+c07,steinbakk+s09,scheel+gr11}), where an attempt is made to perform model checking informative about
different levels of a hierarchical prior.   A discussion of 
checking for conflicting sources of information in a Bayesian model
in a broad sense is given in \cite{presanis+osd13}.   

There has been much discussion of the connection 
between prior-data conflict and imprecise probability models.  
For example, \cite{walter+c16} consider the phenomenon of prior-data conflict
insensitivity for exponential family models and a precise conjugate prior, 
and suggest imprecise probability methods where the range of posterior
inferences is reflective of prior-data conflicts or prior-data agreement.
For earlier related work see
\cite{pericchi+w91}, \cite{coolen94}, \cite{walter+a09} and 
\cite{walter+a10} for example.
This work is insightful about how
various imprecise probability formulations deal with conflicts in simple settings
where computations are tractable, such as for exponential families. 
However, to the best of our knowledge, existing approaches have not been
extended to prior-data conflict checking for density ratio classes, or to models
with intractable likelihood.  We consider
this now.

\subsection{Conventional Bayesian prior-data conflict checks for the bounds}

A first simple approach to conflict checking for density ratio classes is
to apply conventional prior-data conflict checks to the priors specified
by the bounds, $\hat{l}(\theta)$ and $\hat{u}(\theta)$.  If the prior class
represents ambiguity in our prior knowledge, it may be felt that this
prior information should be consistent with the information in the likelihood, 
and that none of the priors in the prior class should have conflict with the
likelihood.  If we take
that view, it can be interesting to 
conduct a conventional Bayesian check on the bounds.  If there is a conflict, 
this suggests there is a problem with the elicitation of the bounds.  
In the next subsection we will consider a different approach, where the goal
is to determine whether every prior in the prior class is in conflict
with the likelihood.  

In implementing a conventional prior-data conflict check
for bounds, we will consider the approach of \cite{evans+m06}.  
Consider a conventional Bayesian analysis with prior $\hat{\eta}(\theta)$.  
\cite{evans+m06}
suggest a prior-data conflict check using the discrepancy 
\begin{align}
  D(T) & =-\log p(T;\hat{\eta}),  \label{emoshonov}
\end{align}
where $T$ is a minimal sufficient statistic.  They calibrate the observed
value $D(T_{\text{obs}})$ of $D(T)$ using 
a prior predictive tail probability
$$P(D(T)\geq D(T_{\text{obs}})),\;\;\;T\sim p(T;\hat{\eta}).$$
The discrepancy used by \cite{evans+m06} is a function of a minimal
sufficient statistic, which is a desirable feature for a prior-data
conflict check.  If
a proposed discrepancy depends on aspects of the data not captured
by a minimal sufficient statistic, then these aspects are irrelevant to the likelihood, 
and can have nothing
to do with whether the likelihood and a prior conflict.  

In SBI, choosing the summary statistic
$S$ as a minimal sufficient statistic is an ideal that is often not attainable, 
but it is desirable to choose an $S$ which
is ``near minimal sufficient''.  Near sufficiency reduces information loss, and
having a statistic of minimal dimension aids computation.  Hence if a good
summary statistic choice has been made for an SBI analysis,
it is natural to use $T=S$ in implementing
the prior-data conflict check of \cite{evans+m06}.  
Computation of discrepancies for the bound checks requires
computation of $p(S;\hat{l})$ and $p(S;\hat{u})$.  These calculations
can be done approximately using \eqref{lowerml}, and 
the calibration tail probability can be estimated by Monte Carlo.  
Considering the check for $\hat{l}(\theta)$,
and substituting a normalizing flow approximations 
$\widetilde{l}(\theta|S)$ and $\widetilde{l}(S|\theta)$ 
for $\hat{l}(\theta|S)$ and $\hat{l}(S|\theta)$ in \eqref{lowerml} we 
obtain an approximate discrepancy
$\widetilde{D}(S)$.   Similarly to \eqref{diffmlbar}, we can 
also consider a geometric average of estimates of $p(S;\hat{l})$ 
from samples $\vartheta_1,\dots, \vartheta_M\sim \widetilde{l}(\theta|S)$ to
obtain an approximate discrepancy $\bar{D}(S)$. A Monte Carlo approximation of a 
Bayesian predictive $p$-value is 
\begin{align*}
\frac{1}{V}\sum_{v=1}^V I\left\{ \bar{D}(S_v)\geq \bar{D}(S_{\text{obs}})\right\},\;\;\;\mbox{for $S_v\sim p(S;\hat{l})$, $v=1,\dots, V$.}
\end{align*}
The check for $\hat{u}(\theta)$ is similar.

A possible problem with the check of \cite{evans+m06} is that the check
is not invariant to the particular form used for the minimal
sufficient statistic $T$.  So, for example, if we make an invertible
transformation of $T$ to $T'$ say, the result of the check may differ.  
Although there are ways to address this issue \citep{evans+j11b}
the checks for density ratio classes in the next
subsection do possess a property of invariance to invertible transformations of 
summary statistics.

\subsection{Prior-data conflict for density ratio classes}

Next we consider checks for whether all the priors in a prior density
ratio class are in conflict with the likelihood.  
If the ``size'' of the prior density ratio class is very large, 
then the checks
we discuss here are unlikely to produce evidence of a conflict.  This is natural, 
because when the degree of prior ambiguity is large, there are many 
different priors compatible with the prior information.  
Hence the checks discussed here are useful mostly
when the elicited density ratio class is highly informative. 

Here we will
consider $r(S)$ defined at \eqref{rScomputable} as the discrepancy for a check,
$$r(S)=r\frac{p(S;\hat{u})}{p(S;\hat{l})}.$$
  Intuitively, 
$r(S)$ will be very large if the prior predictive density value for
$\hat{u}$ is much larger than for $\hat{l}$.  This often happens 
in the case of a prior-data conflict, where prior predictive density values
are very sensitive to the prior, and the usually more diffuse 
$\hat{u}$ places mass closer
to parameter values receiving support from the likelihood. 
As mentioned
earlier in our discussion of example 1, if $\hat{l}$ is less diffuse
than $\hat{u}$, or $\hat{l}(\theta)$ has heavier tails than $\hat{u}(\theta)$
then we would not obtain valid lower and upper bounds with $l(\theta)\leq u(\theta)$ if $\theta$ is unrestricted.
We will compare $r(S_{\text{obs}})$  to what is expected under the model 
to check for conflict.   
It is possible to have density ratio classes where $\hat{l}(\theta)=\hat{u}(\theta)$, and
in this case $r(S)$ is not a suitable discrepancy to use for conflict checking, since
$r(S)$ is then a constant not depending on the data.  Alternative discrepancies that
could be used in such cases are discussed below.

We calibrate our check using the upper probability of the event 
$\{r(S)\geq r(S_{\text{obs}})\}$ for
the prior predictive density ratio class $\psi_{l(S),u(S)}$,
\begin{align}
  \overline{P}(r(S)\geq r(S_{\text{obs}}).  \label{calibration}
\end{align}
As discussed in Section 2.4, the density ratio class $\psi_{l(S),u(S)}$ is larger than 
the set of predictive densities $\{p(S;\hat{\pi}):\hat{\pi}\in \psi_{l(\theta),u(\theta)}\}$, 
leading to conservative Bayesian predictive $p$-values.  
The upper probability $\overline{P}(r(S)\geq r(S_{\text{obs}})$ is  
\begin{align}
  \overline{P}(r(S)\geq r(S_{\text{obs}})) & = 
    \frac{\int_{\{r(S)\geq r(S_{\text{obs}})\}} p(S;\hat{u})\,dS}
    {\int_{\{r(S)\geq r(S_{\text{obs}})\}} p(S;\hat{u})\,dS
    + r^{-1}\int_{\{r(S)< r(S_{\text{obs}})\}} p(S;\hat{l})\,dS}. \label{upper-tail-prob}
\end{align} 
 We use
$\bar{r}(S)$ at \eqref{diffmlbar} to approximate $r(S)$,
and we can approximate the probabilities given by the integrals in 
\eqref{upper-tail-prob} from samples $S_v^u\sim p(S;\hat{u})$, $v=1,\dots, V$
and $S_v^l\sim p(S;\hat{l})$, $v=1,\dots, V$.

The conflict checks considered here are related to the checks
of \cite{evans+m06} with $T=S$, since both are based on prior predictive densities
for $S$.  The statistic
$r(S)$ looks at a ratio of prior predictive density values at $S$ for
$\hat{u}(\theta)$ and $\hat{l}(\theta)$. 
For our robust Bayes conflict check, if we make an
invertible transformation of $S$ to another statistic $S'$, it is easy to see that $r(S)=r(S')$, because $r(S')$ is a ratio of densities where the Jacobian term for the transformation cancels
out.  As mentioned above, if we choose $\hat{u}(\theta)=\hat{l}(\theta)$ in defining
the density ratio class, $r(S)$ cannot be used as a discrepancy for checking for conflict.  
In that case, we could use use the statistic of Evans and Moshonov \eqref{emoshonov}
with $T=S$ and $\hat{\eta}=\hat{u}$, or $\hat{\eta}=\hat{l}$.  Then we can calibrate the
check of the density ratio class with $\overline{P}(D(S)\geq D(S_{\text{obs}}))$.  
As $r\rightarrow 1$, and the density ratio class shrinks towards a single precise prior, 
we would recover the check of \cite{evans+m06} for this prior.

\subsection{Checking for conflicts between summary statistics}

In \cite{mao+wne22}, the authors consider a way of checking for conflict
between different components of a summary statistic vector in a conventional
SBI analysis with summary statistics.  They 
consider dividing $S$ into subvectors $S=(S_A^\top,S_B^\top)^\top$, with
observed value $S_{\text{obs}}=(S_{\text{obs},A}^\top,S_{\text{obs},B}^\top)^\top$, and writing
$$p(\theta|S_{\text{obs}})\propto p(\theta|S_{\text{obs},A})p(S_{\text{obs},B}|S_{\text{obs},A},\theta),$$
we can consider $p(\theta|S_{\text{obs},A})$ as the prior after observing 
$S_A=S_{\text{obs,A}}$ to be updated by the likelihood term $p(S_{\text{obs},B}|S_{\text{obs},A},\theta)$.  If we have checked the adequacy of the model
$p(S_A|\theta)$ for $S_A$, and if there is no prior data conflict between the prior
$p(\theta)$ and $p(S_A|\theta)$, then conflict between
$p(\theta|S_{\text{obs},A})$ and $p(S_{\text{obs},B}|S_{\text{obs},A},\theta)$
indicates conflict between the different subvectors of the
summary statistics.  To get some intuition, a conflict here would suggest
that the values of $\theta$ that give a good fit to $S_{\text{obs},A}$ are not values that
give a good fit to $S_{\text{obs},B}$, which may indicate model misspecification.
The definition of the check of \cite{mao+wne22} is not invariant 
to swapping $S_A$ and $S_B$ in the definition;  the order matters.
If $S_A$ were a sufficient statistic, this check would not be
useful, since in that case $p(S_B|S_{\text{obs},A},\theta)$ does not depend on 
$\theta$, and hence this likelihood term cannot conflict with $p(\theta|S_{\text{obs},A})$
no matter what $S_B$ is observed.  If we have partitioned $S$ into two subvectors
and it is not clear how to order them for the check, we suggest to do the
check for both orderings.  

It is possible to construct a similar check for conflict between summary
statistics for density ratio classes.  
Following the discussion of Section 5.3, the prior-data conflict discrepancy for the case where $S_A$ is observed
but before updating by $S_B$, is
\begin{align}
  r(S_B|S_{\text{obs},A}) & = r(S_{\text{obs},A})\frac{p(S_B|S_{\text{obs},A};\hat{u})}{p(S_B|S_{\text{obs},A};\hat{l})},  \label{discrepancy-ss}
\end{align}
where 
\begin{align}
  p(S_B|S_{\text{obs},A},\hat{u}) & := \frac{p(S_{\text{obs},A},S_B;\hat{u})}{p(S_{\text{obs},A};\hat{u})},  \label{condpredu}
\end{align}
\begin{align}
  p(S_B|S_{\text{obs},A},\hat{l}) & := \frac{p(S_{\text{obs},A},S_B;\hat{l})}{p(S_{\text{obs},A}.\hat{l})},  \label{condpredl}
\end{align}
and
\begin{align}
 r(S_{\text{obs},A})=r\frac{p(S_{\text{obs},A};\hat{u})}{p(S_{\text{obs},A};\hat{l})}
 \label{rsobsa} 
\end{align}
Write $A$ for the event 
$A=\{r(S_B|S_{\text{obs},A})\geq r(S_{\text{obs},B}|S_{\text{obs},A})\}$.  
A calibration tail probability for the discrepancy \eqref{discrepancy-ss} is
\begin{align}
  \overline{P}(A) 
    & = \frac{\int_A p(S_B|S_{\text{obs},A};\hat{u})\,dS_B}
    {\int_A p(S_B|S_{\text{obs},A};\hat{u})\,dS_B
    + r(S_{\text{obs},A})^{-1} \int_{A^c} p(S_B|S_{\text{obs},A};\hat{l})\,dS_B}. \label{upper-tail-prob-ss}
\end{align} 
Further details about approximation of the discrepancy and 
computation are given in Appendix C.

This check may not be very helpful
in the case where the prior density ratio class is large, leading to
very conservative results.  
For checking conflicts between summary statistics, we find it more useful 
to conduct the corresponding checks based on the bounds, and we discuss this now.  
For checking for conflict between summaries, which is a check of the likelihood, 
it is not necessary to consider prior ambiguity, since the prior has nothing
to do with correct specification of the likelihood.

\subsection{Checking for conflicts between summary statistics based on the bounds}

In the check for conflict between summary statistics based on the bounds, we can use
check of \cite{evans+m06}.  If it is observed that
$S_A=S_{\text{obs},A}$, then the discrepancy for the check based on e.g. the lower bound, is
\begin{align*}
D(S_B|S_{\text{obs},A})&=-\log p(S_B|S_{\text{obs},A};\hat{l})
=-\log p(S_{\text{obs},A},S_B;\hat{l})+\log p(S_{\text{obs},A};\hat{l}).
\end{align*}
Estimating the prior predictive density values in the expression on the right can
be done using the methods of Section 4
to obtain an approximate discrepancy 
$\bar{D}(S_B|S_{\text{obs},A})$.  
 To be more explicit, writing 
$\vartheta_1,\dots, \vartheta_M\sim \widetilde{l}(\theta|S_{\text{obs},A})$,
from \eqref{lowerml} 
we can approximate $\log p(S_{\text{obs},A};\hat{l})$ by 
$$\frac{1}{M}\sum_{m=1}^M \left\{\log \hat{l}(\vartheta_m)+\log \widetilde{l}(S_{\text{obs},A}|\vartheta_m)-
\log \widetilde{l}(\vartheta_m|S_{\text{obs},A})\right\},$$
where $\widetilde{l}(S_{\text{obs},A}\mid \vartheta_m)$ is the
normalizing flow likelihood approximation for $p(S_{\text{obs},A}\mid \vartheta_m)$. 
A similar approximation can be made for $\log p(S_{\text{obs},A},S_B;\hat{l})$.

A Monte Carlo approximation of a 
Bayesian predictive $p$-value is obtained as 
\begin{align*}
&\frac{1}{V}\sum_{v=1}^V I\left\{ \bar{D}(S_{v,B}|S_{\text{obs},A})\geq \bar{D}(S_{\text{obs},B}|S_{\text{obs},A})\right\},\;\;\;\mbox{for $S_{v,B}\sim p(S_B|S_{\text{obs},A};\hat{l})$}.
\end{align*}
The check for for the upper bound
of the posterior density ratio class given $S_A=S_{\text{obs},A}$ is similar
to the one for the lower bound.

\section{Examples}\label{sec:example}

We consider three examples.  The first example is a one-dimensional normal
example where all calculations
are done analytically.  The second example is a Poisson example in
one dimension with a non-conjugate prior 
where calculations cannot be done analytically, and it is helpful
for understanding amortized SBI computations for density ratio classes 
and their application to conflict
checking in a simple setting.  The third example is more complex.  
It considers an agent-based model for movement of a species of toads.  The likelihood
is intractable, and we fit the model using a 24-dimensional summary statistic. 
A further time series example is discussed in Appendix D.  

In the examples in Sections 6.2 and 6.3 below, we summarize the results
of the robust Bayesian analysis by plotting the lower and upper bound
functions for the marginal density ratio classes.  The meaning of this
is as follows:  for a certain parameter, say $\theta_1$, if we draw
a curve between the lower and upper bound for the posterior class for $\theta_1$, 
then the normalized version of that is a marginal posterior density 
for $\theta_1$ that arises for some prior in the joint prior density
ratio class.  So the width of the bounds tell us about the potential 
consequences of ignoring the prior ambiuguity and choosing just a single prior
for inference about $\theta_1$.
Density ratio classes also allow us to compute lower and upper probabilities, 
and we make use of this in our predictive checks in Section 6.1.

\subsection{A normal example}

We discuss a simple normal example first, in which 
calculations can be done analytically.   The example comes from
the literature on Bayesian modular inference \citep{liu+bb09}.
There are two data sources, $z=(z_1,\dots, z_{n_1})^\top$ and 
$w=(w_1,\dots, w_{n_2})^\top$.  There are scalar parameters $\varphi$ and $\eta$.  
Given $\varphi$, the elements of $z$ are conditionally independent, 
$z_i|\varphi\sim N(\varphi,1)$, $i=1,\dots, n_1$.  
Give $\varphi$ and $\eta$, the elements of $w$ are conditionally independent,
$w_i|\varphi,\eta\sim N(\varphi+\eta,1)$, $i=1,\dots, n_2$.  
The parameter of interest is $\varphi$, 
and $\eta$ is a bias parameter which affects the data $w$.  
\cite{liu+bb09} consider the setting where $n_1$ is small, and $n_2$ is much larger.  
So $\varphi$ can be estimated using the data $z$ only.  However, 
one might think that since $n_1$ is small, using the biased data $w$ might improve the inference about $\varphi$ substantially.  
\cite{liu+bb09} consider a single prior where $\varphi$ and $\eta$ are independent
{\it a priori} with $\varphi\sim N(0,\delta_1^{-1})$,
$\eta\sim N(0,\delta_2^{-1})$ where $\delta_1,\delta_2>0$ are known precision parameters.
A flat prior on $\varphi$ can be obtained if $\delta_1\rightarrow 0$, and if 
$\delta_2$ is large, this would express confidence that the bias is small.  
\cite{liu+bb09} show that using the biased data $w$ in addition to $z$ is
not helpful for inference about $\varphi$, with limited improvement for small
bias, and very poor inference for large bias.

We consider the checks developed in Sections 5.4 and 5.5 for conflict between summary
statistics for this problem. 
A minimal sufficient statistic for $\theta=(\varphi,\eta)$ is 
$S=(\bar{z},\bar{w})$, where $\bar{z}$ and $\bar{w}$ are 
the sample means of $z$ and $w$
respectively.  
In the notation of Section 5.4, we use $S_A=\bar{z}$ and $S_B=\bar{w}$.  
After observing $\bar{z}$, we can think of the posterior 
given $\bar{z}$ as a prior used for updating by
$S_B=\bar{w}$, and check its consistency with the likelihood term for $\bar{w}$.  
A conflict could indicate inconsistency between the prior and likelihood, 
or a problem with the likelihood;  
here we will consider the situation where the problem is a prior-data conflict, with 
a large bias $\eta$ that conflicts with prior information expressing belief in a 
small bias. 

The lower and upper bound functions $l(\theta)$ and $u(\theta)$ are defined as follows.  
Write $\phi(x;\mu,\sigma^2)$ for the normal density in $x$ with mean $\mu$ and
variance $\sigma^2$.  
Without loss of generality we take the upper bound function to be normalized, 
$$u(\theta)=\hat{u}(\theta)=\phi(\varphi;0,\delta_1^{-1})\times \phi(\eta;0,\delta_2^{-1})$$
and $l(\theta)=r^{-1} \hat{l}(\theta)$ where
$$\hat{l}(\theta)=\phi(\varphi;0,\delta_1^{-1})\times \phi(\eta;0,k\delta_2^{-1}).$$
We set $k$ to be $0.9$ and then multiply $\hat{l}(\theta)$ by $r^{-1}=0.9$ 
which ensures that $l(\theta)$ is a lower bound.  
$\hat{u}(\theta)$ is a joint prior of similar form to the one
considered in \cite{liu+bb09}, and $\hat{l}(\theta)$ 
is similar but with a smaller prior variance for the bias parameter $\eta$.  

Write $\bar{z}_{\text{obs}}$ for the observed value of $\bar{z}$.  
Again using the notation of Section 5.4, routine manipulations show that 
\begin{align}
 &p(S_B|S_{\text{obs},A};\hat{u})  = p(\bar{w}|\bar{z}_{\text{obs}};\hat{u}) =
\phi\left(\bar{w};\frac{n_1}{n_1+\delta_1} \bar{z}_{\text{obs}},\frac{1}{n_1+\delta_1}+\frac{1}{\delta_2}+\frac{1}{n_2}\right),  \label{pred-upper}
\end{align}
and
\begin{align}
 &p(S_B|S_{\text{obs},A};\hat{l})  = p(\bar{w}|\bar{z}_{\text{obs}};\hat{l}) =
\phi\left(\bar{w};\frac{n_1}{n_1+\delta_1} \bar{z}_{\text{obs}},\frac{1}{n_1+\delta_1}+\frac{k}{\delta_2}+\frac{1}{n_2}\right).  \label{pred-lower}
\end{align}
Let's first consider the checks of Section 5.5 for the bounds using the approach
of \cite{evans+m06}.  The discrepancy for the check is $-\log p(\bar{w}|\bar{z}_{\text{obs}},\hat{l})$ for the lower bound and 
$-\log p(\bar{w}|\bar{z}_{\text{obs}},\hat{u})$ for the upper bound.  If we take logs 
in \eqref{pred-upper} and \eqref{pred-lower} and
ignore irrelevant constants we get equivalent discrepancies, which are the same
for both cases and equal to
\begin{align}
\left(\bar{w}-\frac{n_1}{n_1+\delta_1}\bar{z}_{\text{obs}}\right)^2. \label{conflict-toy}
\end{align}  
Let 
\begin{align*}
A=\Bigg\{&\left (\bar{w}-\frac{n_1}{n_1+\delta_1}\bar{z}_{\text{obs}}\right)^2
\geq \left(\bar{w}_{\text{obs}}-\frac{n_1}{n_1+\delta_1}\bar{z}_{\text{obs}}\right)^2 \Bigg\}.
\end{align*}
The probabilities of $A$ under \eqref{pred-upper} and \eqref{pred-lower} respectively 
are
$$P_{A,u}=P\left(W\geq \frac{\left(\bar{w}_{\text{obs}}-\frac{n_1}{n_1+\delta_1}\bar{z}_{\text{obs}}\right)^2}
{\frac{1}{n_1+\delta_1}+\frac{1}{\delta_2}+\frac{1}{n_2}}\right)\hspace{5mm}
P_{A,l}=P\left(W\geq \frac{\left(\bar{w}_{\text{obs}}-\frac{n_1}{n_1+\delta_1}\bar{z}_{\text{obs}}\right)^2}
{\frac{1}{n_1+\delta_1}+\frac{k}{\delta_2}+\frac{1}{n_2}}\right),$$
where $W\sim \chi^2_1$.  $P_{A,u}$ and $P_{A,l}$ are the exact calibration tail
probabilities 
discussed in Section 5.5 for the priors $\hat{u}$ and $\hat{l}$ respectively.  

We can also consider a check of the whole density ratio class (Section 5.4 and Appendix C).  
Since the distribution of $z$ depends only on $\varphi$, and since the $\varphi$ marginal
prior is the same for $\hat{l}(\theta)$ and $\hat{u}(\theta)$, we have $r(S_{\text{obs},A})$ given by \eqref{rsobsa} is $r$, 
and the discrepancy for our conflict check \eqref{discrepancy-ss} is 

\begin{align*}
&r(S_{\text{obs},A})\frac{p(S_B|S_{\text{obs},A};\hat{u})}
{p(S_B|S_{\text{obs},A};\hat{l})}
=r\frac{\phi\left(\bar{w};\frac{n_1}{n_1+\delta_1} \bar{z}_{\text{obs}},\frac{1}{n_1+\delta_1}+\frac{1}{\delta_2}+\frac{1}{n_2}\right)}
{\phi\left(\bar{w};\frac{n_1}{n_1+\delta_1} \bar{z}_{\text{obs}},\frac{1}{n_1+\delta_1}+\frac{k}{\delta_2}+\frac{1}{n_2}\right)}.
\end{align*}

Again an equivalent discrepancy is obtained by taking logs and
ignoring irrelevant constants;  provided that $k\neq 1$ so that $\hat{l}(\theta)\neq \hat{u}(\theta)$, we obtain once more
the discrepancy \eqref{conflict-toy}.    
The calibration (upper) probability \eqref{upper-tail-prob-ss} is
\begin{align}
 & \frac{P_{A,u}}{P_{A,u}+r^{-1}(1-P_{A,l})}.  \label{calibration2}
\end{align}

\begin{figure*}[htbp]
    \centering
    \includegraphics[width=0.9\textwidth]{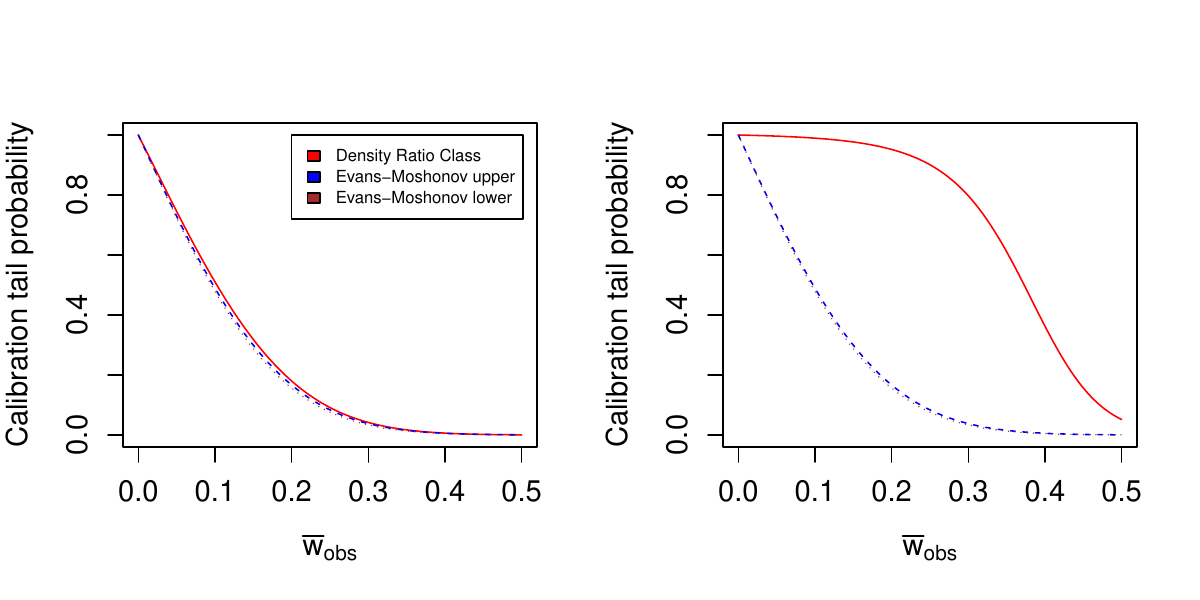}
\caption[]{\label{toynormal} Calibration tail probabilities for density ratio
class and the check of Evans and Moshonov for the bounds for toy normal example
The left graph is for $r=1/0.9$ and the right for $r=100$  The 
curves for the checks of Evans and Moshonov are indistinguishable. } 
\end{figure*}

Consider $n_1=100$, $n_2=1000$, $\delta_1=1$, $\delta_2=100$, $k=0.9$, $r^{-1}=0.9$, 
and $\bar{z}_{\text{obs}}=0$.  Figure \ref{toynormal} (left) shows the calibration probability
\eqref{calibration2} and the probabilities $P_{A,u}$ and $P_{A,l}$ as a function of $\bar{w}_{\text{obs}}$.   The probabilities $P_{A,u}$ and $P_{A,l}$ are
indistinguishable in the Figure, but if $k$ were chosen to be smaller, so that the shape
of $\hat{l}(\theta)$ and $\hat{u}(\theta)$ are very different, then 
they will be distinct.  As the difference between $\bar{z}_{\text{obs}}$ and 
$\bar{w}_{\text{obs}}$ increases, 
the calibration tail probability tends to zero, giving evidence of conflict.
Figure \ref{toynormal} (right) shows a similar situation but with $r=100$.  
As expected it is much less likely that a conflict will be encountered in the
check for the density ratio class because the class of priors is much wider.

\subsection{Poisson example}

The following example was first discussed in 
\cite{sisson+fb18intro}, where it was used 
to demonstrate some difficulties with na\"{i}ve approaches to summary
statistic choice in likelihood-free inference.  We will use it
to illustrate amortized SBI computations for conflict checking with
density ratio classes in a simple one-dimensional case.

In this example we have 5 observations, written as
$y = (y_1, \dots , y_5)^\top$, and the observed value is 
$y_{\text{obs}}=(0, 0, 0, 0, 5)^\top$.  The assumed model is 
$y_i\mid\theta\stackrel{iid}{\sim} \text{Poisson}(\theta)$, $i=1,\dots,5$.  We consider a density ratio class of prior densities for $\theta$ 
with $l(\theta)=(1/3)\hat{l}(\theta)$ where $\hat{l}(\theta)$ 
is a lognormal density
with parameters $\mu=0.25+1/16$, $\sigma=1/4$, and
$u(\theta)=\hat{u}(\theta)$ a lognormal density with
$\mu=0.25+1/4$, $\sigma=1/2$.  
The ratio of integrated upper bound to lower bound is $r=3$. 
Plots of $l(\theta)$ and $u(\theta)$ are shown in Appendix E of the supplementary material.  
Write $S_1(y)=(1/5)\sum_{i=1}^5 y_i$ for the sample mean $\bar{y}$, and $S_2(y)=(1/4)\sum_{i=1}^5 (y_i-\bar{y})^2$ for the sample variance $s^2$.  
Since in the Poisson model the mean and variance are equal to $\theta$, 
$S_1(y)$ and $S_2(y)$ are both sample estimates of $\theta$.  
The observed values of these summary statistics are
$S_1(y_{\text{obs}})=1$ and $S_2(y_{\text{obs}})=5$, and they are very different, 
suggesting that their values conflict and that a model that is over-dispersed with respect
to the Poisson, such as a negative binomial, could be better.  However, in the case
of a likelihood-free analysis using only the summary statistic
$S_1(y)$, or only the summary statistic $S_2(y)$, it is possible
to match the observed summary statistic with the Poisson model.  The summary
statistics are discrete here, but we use continuous
approximations for the sample mean and variance summaries.  

We perform three robust Bayesian analyses
where the summary statistics consist of 1) $S_1(y)$ only 2) $S_1(y)$ and $S_2(y)$ and 3) $S_2(y)$ only.  Since
$S_1(y)=\bar{y}$ is a sufficient statistic for the Poisson model, 
$\hat{u}(\theta|S_{\text{obs}})$ and $\hat{l}(\theta|S_{\text{obs}})$ are the same
for summary statistic choices 1) and 2), but it is interesting to see whether our computational
methods give the same answer for these cases.  For summary 
statistic choice 2), as mentioned above, the observed summary statistic components 
are in conflict and we use this case to demonstrate our conflict checks.  
  
\begin{figure*}[htbp]
	\centering
	\subfloat{
		\begin{minipage}[t]{0.3\linewidth}
			\centering
			\includegraphics[width=1.55in]{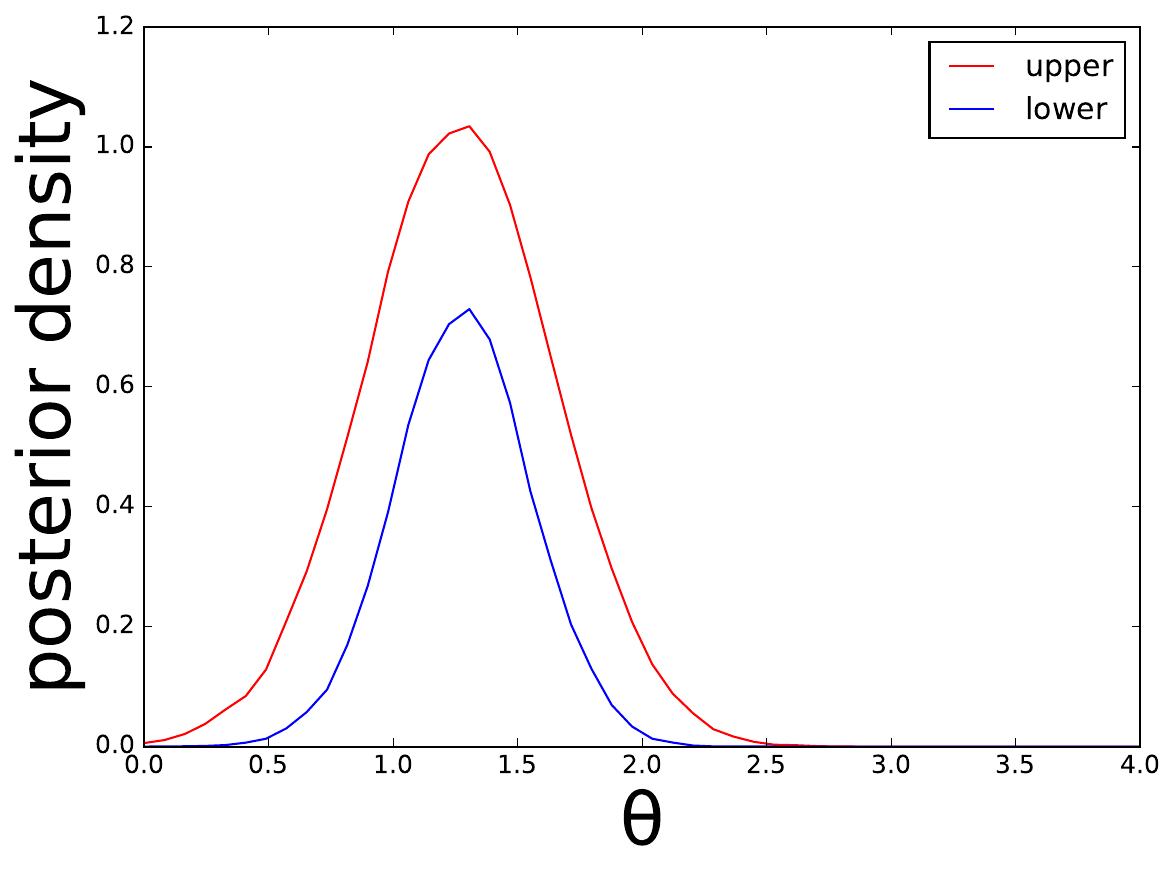}	
             \subcaption{$\bar{y}$}	
		\end{minipage}%
        }%
	\subfloat{
		\begin{minipage}[t]{0.3\linewidth}
			\centering
			\includegraphics[width=1.55in]{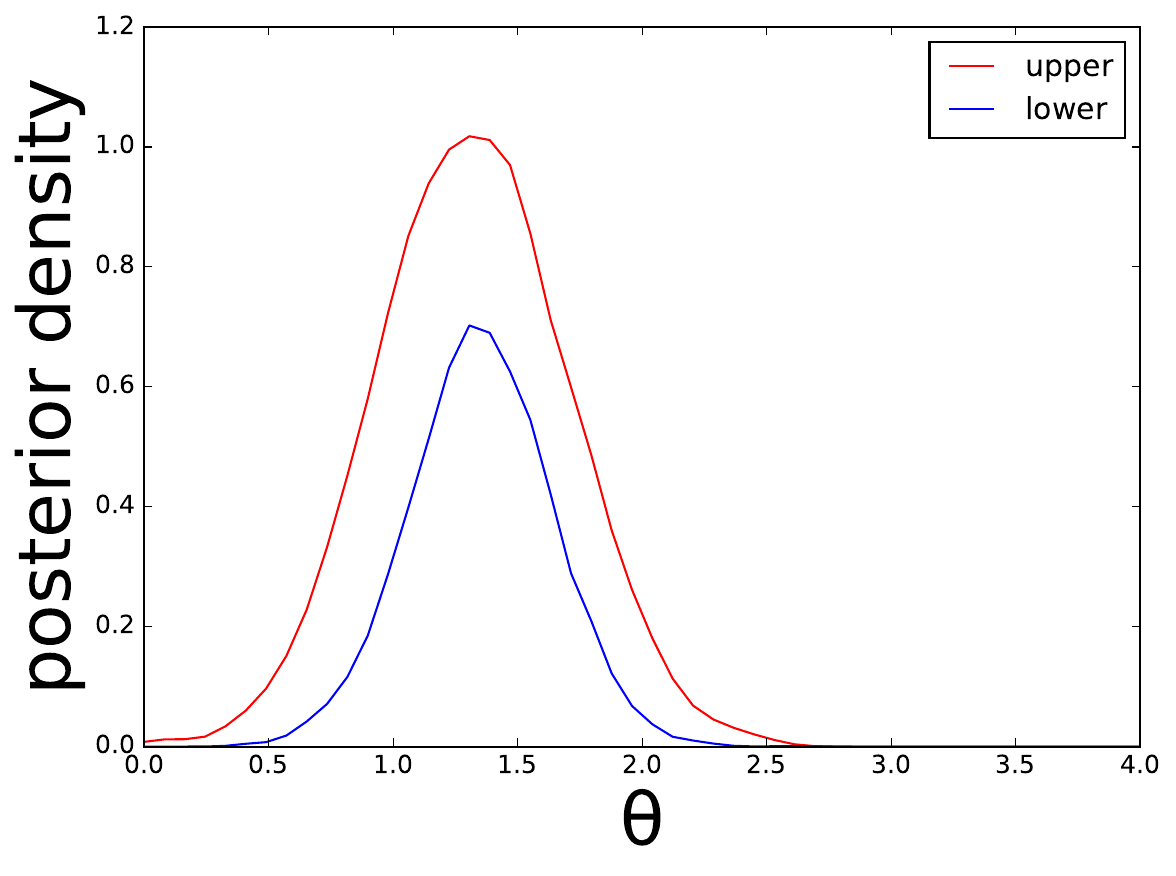}
		 \subcaption{($\bar{y},s^2$)}	
		\end{minipage}%
	}%
	\subfloat{
		\begin{minipage}[t]{0.3\linewidth}
			\centering
			\includegraphics[width=1.55in]{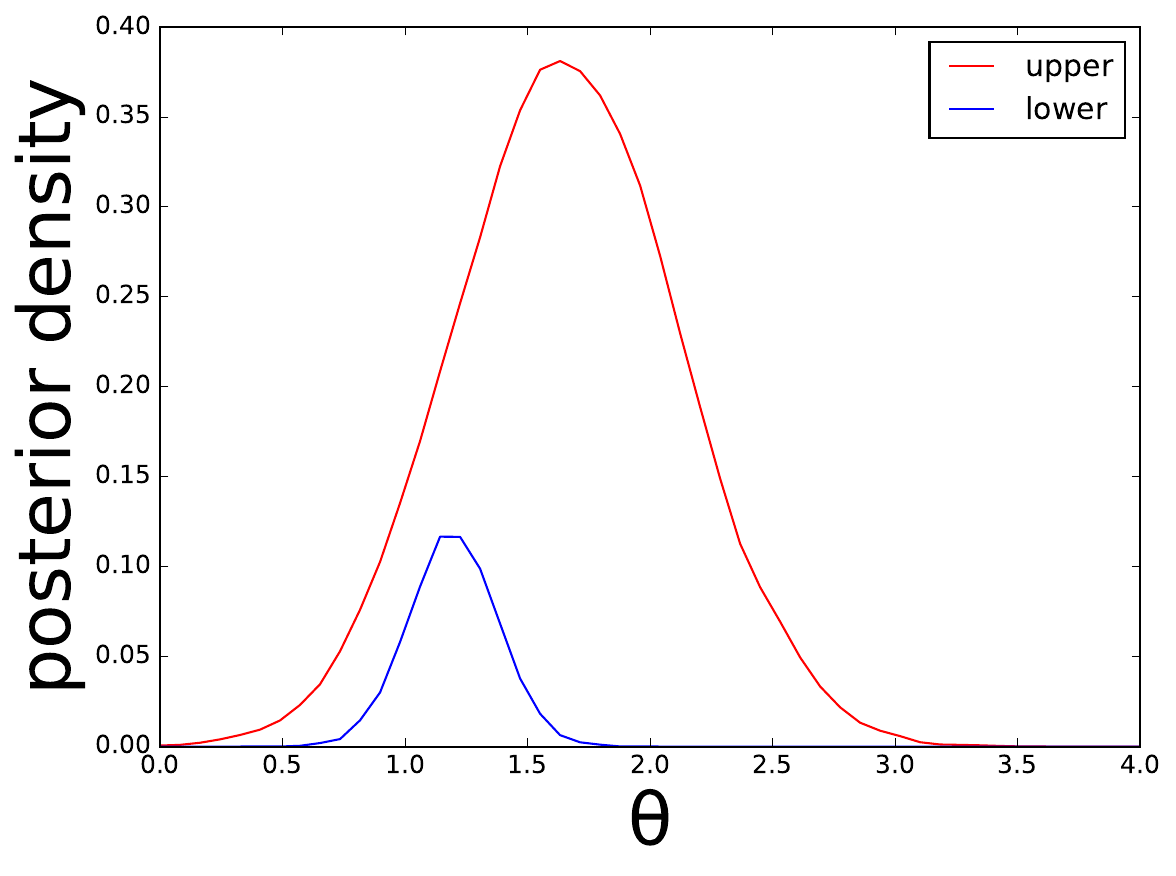}
		\subcaption{$s^2$}	
		\end{minipage}
        }%
	\centering
	\caption[]{\label{toyexample}Estimated upper bound and lower bound for posterior densities via robust Bayes with different summary statistic choices.
	}
\end{figure*}

From simulated data under $\hat{l}(\theta)$ and $\hat{u}(\theta)$
we approximate lower and upper bound functions for the 
posterior density for the 
three different summary statistic choices using the methods of Section 4. The upper bound 
is normalized to be a density function.
Figure \ref{toyexample} shows the results.  For case (a) 
of the figure where the summary statistic 
is the sample mean, the likelihood information is not in 
conflict with the prior information.  We can see that the 
lower and upper bound functions are close. 
For case (b) of the figure where the summary statistic is the 
mean and variance, the lower and upper bound functions 
are close too, and the result is nearly identical to case (a), as expected due
to the sufficiency of the sample mean.
On the other hand, scenario (c) represents a situation where 
the likelihood is less consistent with the prior information, and the prior ambiguity
leads to greater posterior ambiguity, because of the greater posterior sensitivity
to the prior when the prior and likelihood conflict.
Calibration plots for the
three different cases to check the accuracy of the amortized inference
computations are shown in Appendix E of the supplement.  We used
the same settings for training in each case, but 
the calibration plots suggest that for 
the upper bound prior $\hat{u}(\theta)$ and $S_2(y)$ 
the posterior estimation is not reliable.

Checking for prior-data conflict between the density ratio class and the
likelihood using the discrepancy in Section 5.2 using calibration \eqref{upper-tail-prob}  
we obtain upper tail probabilities for the three summary statistic choices of
$\color{blue}{0.998}$ for case 1, $\color{blue}{0.956}$ for case 2, and $\color{blue}{0.164}$ for case 3.  This indicates some conflict
between the likelihood and prior density ratio class in case 3, where the data information
consists of just the sample variance summary which is inconsistent with 
the prior information.  
We also check for conflict between the two summary statistic
components as described in Sections 5.4 and 5.5 with $S_A=s^2$ and $S_B=\bar{y}$.  
As discussed in Section 5.4, we should not choose $S_A$ to be the sufficient
statistic $\bar{y}$.  It is also not appropriate to do our conflict check 
between summaries 
if the prior density ratio class or its bounds are in conflict with the information in $S_A$, and 
we consider new upper and lower bound functions that are not in conflict with
$s^2=5$ before performing the check.  We let
$l(\theta)=(1/2)\hat{l}(\theta)$ where $\hat{l}(\theta)$ 
is a lognormal density
with parameters $\mu=1.09 + 1/9$, $\sigma=1/3$, and
$u(\theta)=\hat{u}(\theta)$ a lognormal density with
$\mu=1.09 + 1/16$, $\sigma=1/4$.  With these choices $\hat{l}(\theta)$ and
$\hat{u}(\theta)$ are roughly peaked at $3$, and there is no conflict 
between $S_A$ and the prior.  
The calibration tail probability for the conflict check
between the summaries for the prior density ratio class (Section 5.4 and Appendix C)
is $\color{blue}{0.793}$, indicating a lack of conflict, due to the conservatism caused by the
prior ambiguity.  
In the check for conflict between summary statistics based on the bounds 
using the method of \cite{evans+m06} (Section 5.5), 
for the lower bound the calibration tail probability is $\color{blue}{0.0030}$ and 
for the upper bound it is $\color{blue}{0.039}$, which indicates conflict 
and the problem with the specification of the model here.  

\subsection{Toad example}\label{toad}

Our next example considers an agent-based model for movement of a species
of toads ({\it Anaxyrus floweri}, or Fowler's toads).  We consider data originally discussed
in \cite{marchand17}, also analyzed in \cite{frazier+d21}.  
The data are modelled using `model 2' in \cite{marchand17}, which has a parameter
$\theta=(\alpha,\delta,p_0)^\top$, where $\alpha$ and $\delta$ are stability and
scale parameters in an alpha-stable distribution for an overnight displacement in 
toad movements, and $p_0$ is a probability for a toad returning to a previously
used refuge.  

The raw data consists of GPS locations for 63 days and 66 toads.  The summary statistic
vector $S$ that we consider 
has dimension $24$, with $S=(S_A^\top,S_B^\top)^\top$, and
$S_A$ and $S_B$ are 12-dimensional summary statistic vectors.  The vector $S_A$
summarizes the displacements for all toads for a one day lag;  the vector $S_B$ summarizes
the displacements for all toads for a two day lag.  The analysis in
\cite{frazier+d21} suggests conflict between the summary statistics at different lags
(i.e. between $S_A$ and $S_B$).  They also consider summary statistics at 
lags of 4 and 8 days, but we do not do this here.  

\begin{figure*}[htbp]
\centering
\includegraphics[width=0.9\textwidth]{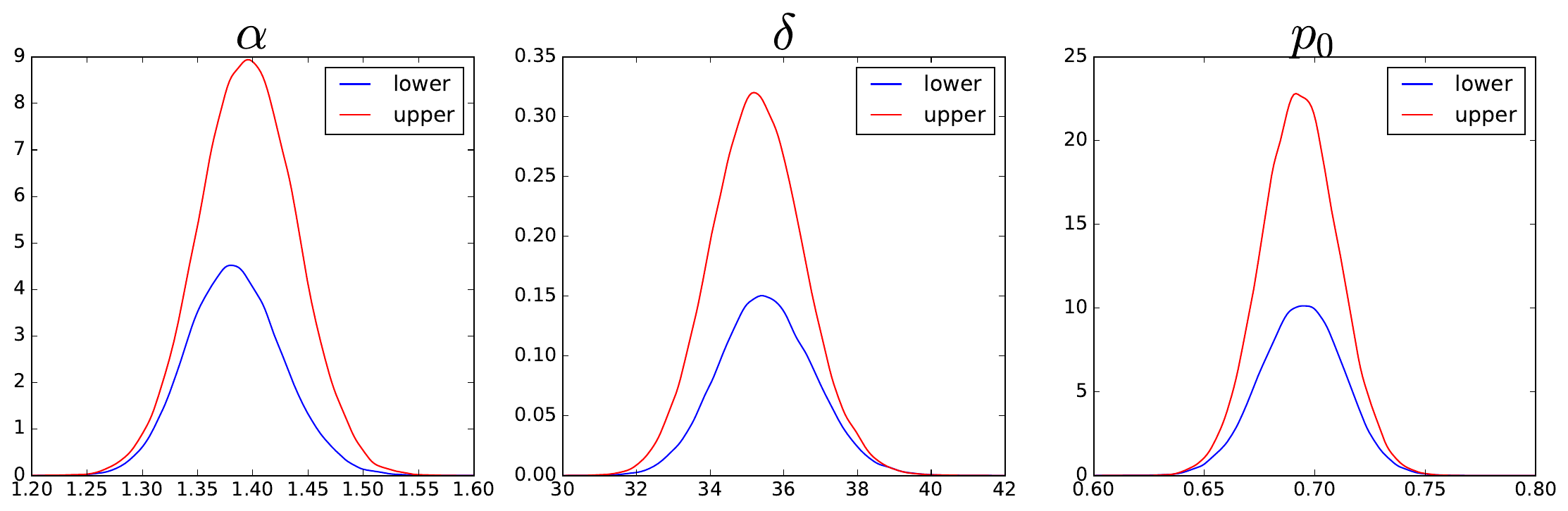}

\vspace{1ex}
\includegraphics[width=0.9\textwidth]{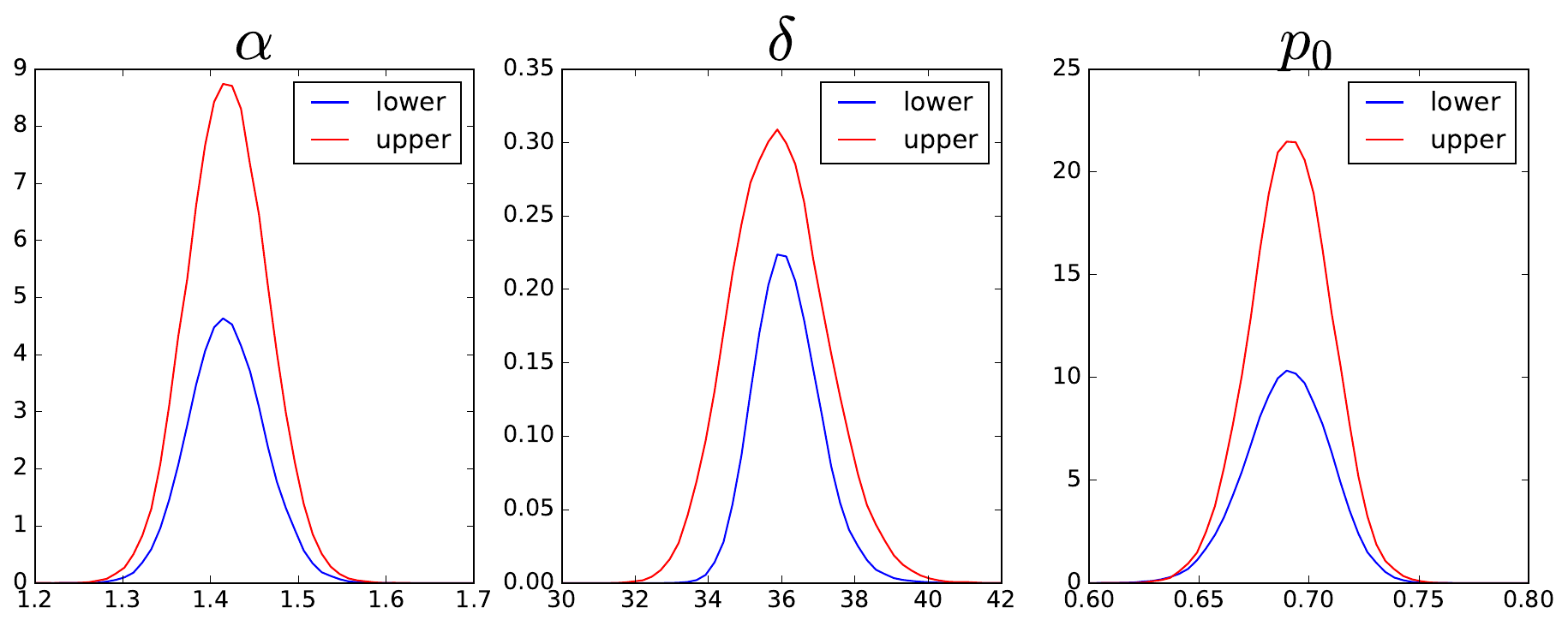}

\vspace{1ex}
\includegraphics[width=0.9\textwidth]{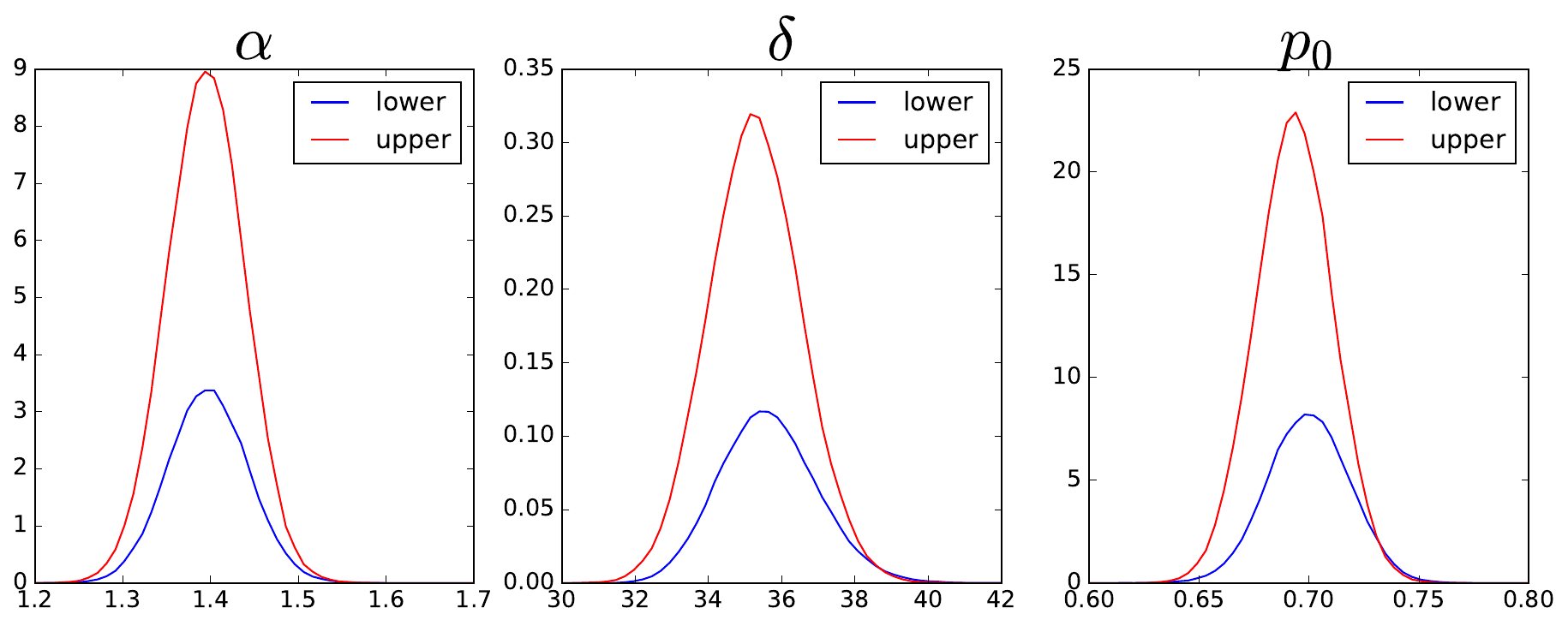}

\caption{ Estimated upper bound and lower bound functions for
	marginal posterior density ratio classes for case 1 (first row), 
	case 2 (second row) and case 3 (third row) for the Toad example.	}
\label{toad-alpha-real}
\end{figure*}

We will consider three different prior density ratio classes. 
In all three cases, 
the upper bound $u(\theta)$ will be a uniform density for
$\theta$ on the range
$\left[ 1, 2\right]\times\left[30, 50 \right]\times\left[0, 0.9 \right]$.
The lower bound functions $l(\theta)$ are different in the three cases.
The first case (hereafter case 1) takes $l(\theta)=(1/3)\hat{l}(\theta)$ and
$\hat{l}(\theta)$ a
uniform density on $\left[ 1.2, 1.8\right]\times\left[30, 50 \right]\times\left[0, 0.9 \right]$.  The second case (hereafter case 2) takes 
$l(\theta)=(1/3)\hat{l}(\theta)$ and
$\hat{l}(\theta)$ a uniform density 
on $\left[ 1, 2\right]\times\left[35, 45 \right]\times\left[0, 0.9 \right]$.
The third case (hereafter case 3) takes 
$l(\theta)=(1/3)\hat{l}(\theta)$ and
$\hat{l}(\theta)$ a uniform density
on $\left[ 1, 2\right]\times\left[30, 50 \right]\times\left[0.1, 0.8 \right]$.
In case 1, $\hat{l}(\theta)$ and $\hat{u}(\theta)$ are densities
where $\theta_1,\theta_2$ and $\theta_3$ are independent, 
with the marginal densities for $\theta_2$ and $\theta_3$ being
the same for $\hat{l}(\theta)$ and $\hat{u}(\theta)$, whereas
the densities for $\theta_1$ are different. Case 2 and case 3 are similar to case 1 with the different densities for $\theta_2$ and $\theta_3$ respectively.
The summary statistic data comes from the BSL R package.
Considering case 1 above, and normalizing the upper bound to be a density function, Figure \ref{toad-alpha-real} shows the estimated lower and upper bound functions $\bar{r}(S_{\text{obs}})^{-1}\widetilde{l}(\theta_j|S_\text{obs})$ and $\widetilde{u}(\theta_j|S_\text{obs})$ for 
the marginal posterior density ratio classes for the three parameters, $j=1,2,3$ for the three
cases for the specified prior density ratio class.  Diagnostic plots examining the reliability
of the amortized inference computation are shown in Appendix A.  

		

We also do the summary statistic conflict checking as described in Section 5.5 
based on the bounds.  
 The calibration tail probabilities for conflict between
the summaries based on the bounds are $<0.001$
for both the upper and lower bounds in all three cases. 
This suggests conflict between the summary statistics at different time lags, 
which was also the conclusion of \cite{frazier+d21}.  

\section{Discussion}

Much recent work on neural SBI methods has focused on achieving robustness
to misspecification of the likelihood component of the model.  Complementing
this work, we consider issues of robustness to the choice of prior, and implementing
robust Bayes methods which avoid the choice of a single prior.  We demonstrate that
recently developed amortized neural SBI methods can be adapted 
to compute robust Bayesian
inferences based on density ratio classes.  Methods of checking for
conflict between a density ratio class and the likelihood, and checking for conflict
between subsets of summary statistics, are also developed.   Conflict checks 
can be based on checking whether all priors in a density ratio class conflict with
likelihood information, or on whether one of the bounds is in conflict, where conventional
Bayesian conflict checking methods can be employed.  In the latter case, the prior
class may contain some reasonable priors and some unreasonable priors, in the
sense of conflicting with the likelihood information.  
It would be possible to combine the methods discussed in this work with any of the
previously suggested methods in the literature 
for robustifying the model through model expansions
to obtain robustness to both likelihood misspecification and prior ambiguity or prior-data
conflicts. A difficulty with all robust Bayes methods is the challenge of 
eliciting the prior class used.  While \cite{rinderknecht+br11} have done some pioneering
work in this direction for density ratio classes, this remains a difficult task in models
for which the parameter is high-dimensional.

\subsection*{Acknowledgements}
David Nott's research was supported by the Ministry of Education, Singapore, under the Academic Research Fund Tier 2 (MOE-T2EP20123-0009).  Evans' research was supported
by a grant from the Natural Sciences and Engineering Research Council of Canada.
Wang Yuyan thanks the developers of the JANA package and Stefan Radev for patiently answering her questions.

\section*{Appendix A:  Previous work on SBI with misspecified likelihood}

One of the earliest works on model misspecification and SBI is
\cite{ratmann+awr09}, who consider model expansions for ABC regarding tolerances
as model parameters.  
\cite{wilkinson13} demonstrates that the posterior approximation in ABC is an exact
posterior by interpreting the kernel and tolerance in ABC procedures 
as specifying a model error.  
\cite{frazier+rr20} examine theoretically the behaviour of standard and regression-adjustment ABC methods
under misspecification, and suggest diagnostics.  
\cite{lewis+ml21} give a general discussion of the value of using insufficient 
summary statistics to robustify Bayesian modelling, and discuss
implementing exact conditioning for certain types of summary statistics 
in linear model settings.  \cite{frazier+d21}
suggest model expansions for the Bayesian synthetic likelihood approach, where 
a parameter is introduced for each summary statistic.  The added parameters 
can be in a mean or variance adjustment, and they discount a
sparse set of incompatible summary statistics which cannot be matched 
under the assumed model.  The approach is useful for understanding the 
nature of misspecification through the summary statistics, which can 
guide model improvement.  \cite{frazier+nd24} demonstrate that the 
standard synthetic likelihood posterior can exhibit non-standard behaviour 
under misspecification.  Numerous authors have warned about the difficulties
arising with neural SBI methods under misspecification 
\citep{cannon+ws22,ward+cbfs22,huang+bsak23,ryan+nfwd24}, with suggested
remedies including model expansions similar to those considered
by \cite{frazier+rr20} in the BSL context, methods for choosing summary statistics where incompatibility is penalized, or addition of noise to summary statistics
in training.  Another relevant approach
to dealing with misspecification is to use generalized Bayesian inference \citep{bissiri+hw16} where
the likelihood is replaced by a pseudo-likelihood derived from a loss function. 
In the case of SBI methods, the loss function is often based on a scoring
rule \citep{giummole+mrv19}.  Recent work on generalized Bayesian 
methods for SBI includes  
\cite{cherief-abdellatif+ba20}, \cite{schmon+ck21}, \cite{matsubara+kbo22}, \cite{dellaporta+kdb22}, \cite{gao+dm23}, \cite{pacchiardi+kd24} and
\cite{weerasignhe+lmf25}.

\section*{Appendix B: Basic concepts for density ratio classes}

\subsection*{Equivalence of definitions of likelihood ratio classes}

Here we consider the equivalence of the two definitions of a density ratio class given
in the main text.  
Write the lower bound function as $l(\theta)$ and the upper bound function as $u(\theta)$,
$0\leq l(\theta)\leq u(\theta)$, with
$$\int l(\theta)\,d\theta>0\;\;\;\text{and}\;\;\;\int u(\theta)\,d\theta<\infty$$
The first definition of a density ratio class is
\begin{align}
  \psi_{l,u} := \left\{\hat{\pi}(\theta)=\frac{\pi(\theta)}{\int \pi(\theta)\,d\theta};
  l(\theta)\leq \pi(\theta)\leq u(\theta)\right\}. 
 \end{align}  
In the case where $l(\theta)>0$ for all $\theta\in\Theta$, an equivalent definition of $\psi_{l,u}$ is
\begin{align}
  \psi_{l,u} & = \left\{\hat{\pi}(\theta): \frac{l(\theta)}{u(\theta')}\leq \frac{\pi(\theta)}{\pi(\theta')}\leq \frac{u(\theta)}{l(\theta')}\mbox{ for all $\theta,\theta'\in \Theta$}\right\}.  
\end{align}

To demonstrate the
equivalence, start with definition \eqref{drclass}.  If 
$\hat{\pi}(\theta)$ belongs to $\psi_{l,u}$ in \eqref{drclass}, then for any given
$\theta,\theta'$, and assuming $l(\theta)>0$ for all $\theta$, we can write 
$$l(\theta)\leq \pi(\theta)\leq u(\theta),\;\;\;1/u(\theta')\leq 1/\pi(\theta')\leq 1/l(\theta').$$
Multiplying terms in the inequalities gives
$$l(\theta)/u(\theta')\leq \pi(\theta)/\pi(\theta')\leq u(\theta)/l(\theta'),$$
and $\hat{\pi}(\theta)$ belongs to $\psi_{l,u}$ in \eqref{drclass2}.  

In the other direction, assume that $\hat{\pi}(\theta)$ belongs to 
$\psi_{l,u}$ given in \eqref{drclass2}, with $l(\theta)>0$ for all $\theta\in\Theta$.  
Then for all $\theta,\theta'\in \Theta$, 
$$\frac{\pi(\theta)}{\pi(\theta')}\leq \frac{u(\theta)}{l(\theta')},$$
and setting $\theta=\theta'$ we can deduce that $l(\theta)\leq u(\theta)$
for all $\theta\in\Theta$, and hence $u(\theta)>0$ for all $\theta\in\Theta$.
Rearranging the above inequality, 
\begin{align}
  \frac{\pi(\theta)}{u(\theta)} & \leq \frac{\pi(\theta')}{l(\theta')}, \label{pioveru}
\end{align}
and
$$c=\sup_{\theta\in\Theta} \frac{\pi(\theta)}{u(\theta)}<\infty,$$
which implies that $\pi(\theta)\leq cu(\theta)$ for all $\theta$.  
Suppose that there exists $\theta'$ such that $\pi(\theta')<cl(\theta')$, which implies that
\begin{align}
  \frac{\pi(\theta')}{l(\theta')}<c.  \label{pioverl}
\end{align}
Then from \eqref{pioveru} and \eqref{pioverl} and for any $\theta$,
$$\frac{\pi(\theta)}{u(\theta)}\leq \frac{\pi(\theta')}{l(\theta')}<c,$$
for all $\theta$, which implies
$$c=\sup_\theta \frac{\pi(\theta)}{u(\theta)}<c,$$
a contradiction.  So we must have $cl(\theta)\leq \pi(\theta)\leq cu(\theta)$, so
that $\hat{\pi}(\theta)$ belongs to $\psi_{l,u}$ defined by \eqref{drclass}.

\subsection*{Derivations for Example 1 in the main text}

The inequality $l(\theta) \leq u(\theta)$ for all $\theta \in \mathbb{R}^p$ is equivalent to $-2\log l(\theta)\geq -2\log u(\theta)$, which simplifies to
\begin{equation}
\sum_{j=1}^p \left[ \frac{(\theta_j - \mu_{u,j})^2}{\tau_{u,j}^2} - \frac{(\theta_j - \mu_{l,j})^2}{\tau_{l,j}^2} \right] \;\leq\; 2\log r - \sum_{j=1}^p \log\frac{\tau_{u,j}^2}{\tau_{l,j}^2} \quad \text{for all } \theta. \label{eq:condition}
\end{equation}
The $j$th term in the sum on the left is quadratic in $\theta_j$ with the coefficient on $\theta_j^2$ being 
$\tau_{u,j}^{-2} - \tau_{l,j}^{-2}$.  We consider three cases:
\begin{itemize}
\item If $\tau_{l,j}^2 < \tau_{u,j}^2$, the coefficient on $\theta_j^2$ is negative and 
the $j$th term on the left-hand side has a finite maximum of $(\mu_{u,j} - \mu_{l,j})^2/(\tau_{u,j}^2 - \tau_{l,j}^2)$.
\item If $\tau_{l,j}^2 = \tau_{u,j}^2$ and $\mu_{l,j} = \mu_{u,j}$, the $j$th term is identically zero in $\theta_j$, and the corresponding term $\log(\tau_{u,j}^2/\tau_{l,j}^2)$ on the right-hand side is also zero, so the $j$th terms in the summations contribute nothing.
\item If either $\tau_{l,j}^2 > \tau_{u,j}^2$ or $\tau_{l,j}^2 = \tau_{u,j}^2$ with $\mu_{l,j} \neq \mu_{u,j}$, the $j$th term in the sum on the left-hand side is unbounded above in $\theta_j$. In this case~\eqref{eq:condition} cannot hold for all $\theta$.
\end{itemize}
Let $J = \{j : \tau_{l,j}^2 < \tau_{u,j}^2\}$. Assuming for each $j \notin J$ that $\tau_{l,j}^2 = \tau_{u,j}^2$ and $\mu_{l,j} = \mu_{u,j}$, the supremum of the left side of~\eqref{eq:condition} over $\theta$ is finite and equals $\sum_{j \in J}(\mu_{u,j} - \mu_{l,j})^2/(\tau_{u,j}^2 - \tau_{l,j}^2)$, and the right-hand side reduces to $2\log r - \sum_{j \in J} \log(\tau_{u,j}^2/\tau_{l,j}^2)$, giving equation (5) in the main text.

Next let's derive the lower and upper bounds for the joint and marginal posterior density ratio classes.  By the closure of density ratio classes under Bayesian updating, the joint posterior class has unnormalized bounds $l(\theta| \bar{y}) = l(\theta) p(\bar{y} \mid \theta)$ and $u(\theta| \bar{y}) = u(\theta) p(\bar{y} \mid \theta)$. Since $\Sigma$ and the prior covariance matrices are all diagonal, the posterior update corresponds to univariate updates coordinate-wise. For each $j$, multiplying $\phi(\theta_j; \mu_{l,j}, \tau_{l,j}^2)$ by $\phi(\bar{y}_j; \theta_j, \sigma_j^2/n)$ and exploiting conjugacy gives
\[
\phi(\theta_j; \mu_{l,j}, \tau_{l,j}^2)\,\phi(\bar{y}_j; \theta_j, \sigma_j^2/n) = Z_{l,j}\,\phi(\theta_j; \mu_{l,j}^\star, \tau_{l,j}^{\star 2}),
\]
with the conjugate updates for $(\mu_{l,j}^\star, \tau_{l,j}^{\star 2})$ as stated in the main text, and multiplicative constant
\[
Z_{l,j} = \phi\bigl(\bar{y}_j; \mu_{l,j}, \tau_{l,j}^2 + \sigma_j^2/n\bigr).
\]
A similar update holds for the upper bound with constant $Z_{u,j} = \phi(\bar{y}_j; \mu_{u,j}, \tau_{u,j}^2 + \sigma_j^2/n)$. Hence
\[
l(\theta| \bar{y}) = r^{-1}\prod_{j=1}^p Z_{l,j}\,\phi(\theta_j; \mu_{l,j}^\star, \tau_{l,j}^{\star 2}), \qquad u(\theta| \bar{y}) = \prod_{j=1}^p Z_{u,j}\,\phi(\theta_j; \mu_{u,j}^\star, \tau_{u,j}^{\star 2}),
\]
which after normalization gives the bounds in equation~(8) of the main text. 
The ratio of integrated upper to integrated lower bound (equation~(9) of the main text) is
\[
r(\bar{y}) = \frac{\int u(\theta| \bar{y})\,d\theta}{\int l(\theta| \bar{y})\,d\theta} = r \cdot \prod_{j=1}^p \frac{Z_{u,j}}{Z_{l,j}}.
\]

By the closure of density ratio classes under marginalization, the marginal class for $\theta_A$ has bound functions $\int l(\theta|\bar{y})\,d\theta_B$ and $\int u(\theta|\bar{y})\,d\theta_B$. Because the posterior bounds in equation~(8) in the main text factorize across coordinates and each univariate normal density integrates to one, the integrals over $\theta_B$ result in the marginal Gaussian bounds given in
equation~(10) in the main text.

\section*{Appendix C:  Checking for conflict between summary statistics using
density ratio classes}

Consider a similar check to Section 5.3, where
now $p(\theta|S_{\text{obs},A})$ is the prior and $p(S_{\text{obs},B}|S_{\text{obs},A},\theta)$ is the likelihood.  Our prior-data conflict discrepancy, for the case where $S_A$ is observed
but before updating by $S_B$, is
\begin{align}
  r(S_B|S_{\text{obs},A}) & = r(S_{\text{obs},A})\frac{p(S_B|S_{\text{obs},A},\hat{u})}{p(S_B|S_{\text{obs},A},\hat{l})},  \label{discrepancy-ss}
\end{align}
where 
\begin{align}
  p(S_B|S_{\text{obs},A},\hat{u}) & := \frac{p(S_{\text{obs},A},S_B;\hat{u})}{p(S_{\text{obs},A};\hat{u})},  \label{condpredu}
\end{align}
and
\begin{align}
  p(S_B|S_{\text{obs},A},\hat{l}) & := \frac{p(S_{\text{obs},A},S_B;\hat{l})}{p(S_{\text{obs},A}.\hat{l})},  \label{condpredl}
\end{align}
Using \eqref{condpredu} and equation (13) in the main text we obtain
\begin{align}
p(S_B|S_{\text{obs},A},\hat{u}) & = \frac{\hat{u}(\theta)p(S_{\text{obs},A},S_B|\theta)}
{\hat{u}(\theta|S_{\text{obs},A},S_B)}\times \frac{\hat{u}(\theta|S_{\text{obs},A})}{\hat{u}(\theta)p(S_{\text{obs},A}|\theta)}, \label{cden1}
\end{align}
and similarly from \eqref{condpredl} and equation (13) in the main text
\begin{align}
  p(S_B|S_{\text{obs},A},\hat{l}) & = \frac{\hat{l}(\theta)p(S_{\text{obs},A},S_B|\theta)}
{\hat{l}(\theta|S_{\text{obs},A},S_B)}\times \frac{\hat{l}(\theta|S_{\text{obs},A})}{\hat{l}(\theta)p(S_{\text{obs},A}|\theta)}.  \label{cden2}
\end{align}
Also, 
\begin{align}
 r(S_{\text{obs},A})=r\frac{p(S_{\text{obs},A};\hat{u})}{p(S_{\text{obs},A};\hat{l})}
=r\frac{\hat{u}(\theta)\hat{l}(\theta|S_{\text{obs},A})}{\hat{l}(\theta)\hat{u}(\theta|S_{\text{obs},A})}, \label{rsobsa} 
\end{align}
for any value of $\theta$, where the last equality above comes from equation (14) in the
main text.
Using the same value of $\theta$ in the expressions
\eqref{cden1}, \eqref{cden2} and \eqref{rsobsa} we obtain for \eqref{discrepancy-ss}
\begin{align}
  r(S_B|S_{\text{obs},A})=r \frac{\hat{u}(\theta)\hat{l}(\theta|S_{\text{obs},A},S_B)}{\hat{l}(\theta)\hat{u}(\theta|S_{\text{obs},A},S_B)}.  \label{discrepancy-ss-comp}
\end{align}
Using normalizing flow approximations 
$\widetilde{u}(\theta|S_{\text{obs},A},S_B)$ and $\widetilde{l}(\theta|S_{\text{obs},A},S_B)$ 
for 
$\hat{u}(\theta|S_{\text{obs},A},S_B)$ and
$\hat{l}(\theta|S_{\text{obs},A},S_B)$ respectively gives an approximation of \eqref{discrepancy-ss-comp}, which we denote by $\widetilde{r}(S_B|S_{\text{obs},A})$.  
 Letting $\vartheta_1,\dots, \vartheta_M\sim \widetilde{u}(\theta|S_{\text{obs},A},S_B)$ we can also consider an estimate
$$\bar{r}(S_B|S_{\text{obs},A}):=r\prod_{m=1}^M \left\{\frac{\hat{u}(\vartheta_m)\hat{l}(\vartheta_m|S_{\text{obs},A},S_B)}{\hat{l}(\vartheta_m)\hat{u}(\vartheta_m|S_{\text{obs},A},S_B)}\right\}^{1/M}$$
The density ratio class
for the posterior density given $S_A=S_{\text{obs},A}$ has
lower and upper bound functions
$$r(S_{\text{obs},A})^{-1}\hat{l}(\theta|S_{\text{obs},A})<\hat{u}(\theta|S_{\text{obs},A}).$$
where $\hat{l}(\theta|S_{\text{obs},A})$ and $\hat{u}(\theta|S_{\text{obs},A})$
are posterior densities for $\theta$ given $S_A=S_{\text{obs},A}$ for 
priors $\hat{l}(\theta)$ and $\hat{u}(\theta)$ respectively.
This density ratio class 
leads to a density ratio class of conditional predictive densities of $S_B$ given
$S_A=S_{\text{obs},A}$, 
$\psi_{l(S_B|S_{\text{obs},A}),u(S_B|S_{\text{obs},A})}$, where
$$l(S_B|S_{\text{obs},A})=\int p(S_B|S_{\text{obs},A},\theta)r(S_{\text{obs},A})^{-1}\hat{l}(\theta|S_{\text{obs},A})\,d\theta$$
and
$$u(S_B|S_{\text{obs},A})=\int p(S_B|S_{\text{obs},A},\theta)\hat{u}(\theta|S_{\text{obs},A})\,d\theta.$$
This density ratio class is larger than the set of predictive
distributions $\{p(S_B|S_{\text{obs},A};\hat{\pi}):\hat{\pi}(\theta) \in 
\psi_{l,u}\}$, 
leading to conservative Bayesian predictive $p$-values when 
calibrating the discrepancy for our conflict check.  

Write $A$ for the event 
$A=\{\bar{r}(S_B|S_{\text{obs},A})\geq \bar{r}(S_{\text{obs},B}|S_{\text{obs},A})\}$. A calibration tail probability for the discrepancy \eqref{discrepancy-ss} is
\begin{align}
  \overline{P}(A) 
    & = \frac{\int_A p(S_B|S_{\text{obs},A};\hat{u})\,dS_B}
    {\int_A p(S_B|S_{\text{obs},A};\hat{u})\,dS_B
    + r(S_{\text{obs},A})^{-1} \int_{A^c} p(S_B|S_{\text{obs},A};\hat{l})\,dS_B}. \label{upper-tail-prob-ss}
\end{align} 
The integrals in the above expression can be approximated by
$$\int_A p(S_B|S_{\text{obs},A};\hat{u}_{S_A})\,dS_B \approx 
\frac{1}{V}\sum_{v=1}^V I\left\{\bar{r}(S_{v,B}^{u,B|A}|S_{\text{obs},A})\geq \bar{r}(S_{\text{obs},B}|S_{\text{obs},A})\right\},$$
for $S_v^{u,B|A}\sim p(S_B|S_{\text{obs},A};\hat{u}_{S_A})$, $v=1,\dots, V$, 
and
$$\int_{A^c} 
p(S_B|S_{\text{obs},A};\hat{l}_{S_A})\,dS_B\approx
\frac{1}{V}\sum_{v=1}^V I\left\{\bar{r}(S_{v,B}^{l,B|A}|S_{\text{obs},A})< \bar{r}(S_{\text{obs},B}|S_{\text{obs},A})\right\},$$
for $S_v^{l,B|A}\sim p(S_B|S_{\text{obs},A};\hat{l}_{S_A})$, $v=1,\dots, V$.
Approximate simulation of replicates from $p(S_B|S_{\text{obs},A};\hat{u})$ 
could be done using many approaches, but a simple approach that we use here
is based on an ABC approximation.  We simulate a large number $L$ of replicates $(\theta^l,S_A^l,S_B^l)\sim \hat{u}(\theta)p(S|\theta)$, and then select from these the $V$
replicates for which $S_A^l$ is closest to $S_{\text{obs},A}$ in some distance, such
as Euclidean distance.  
Simulation of replicates from $p(S_B|S_{\text{obs},A};\hat{l})$ is
done similarly.

\subsection*{Appendix D:   Ricker model}\label{Ricker}

We consider an additional time series application involving 
the Ricker model \citep{ricker1954},
a simple model for population sizes in ecology.
Let $N_t$, $t=1,\dots, T$,  
be population sizes, with observations
$d_t \sim$ Poisson($\phi N_t$), where $\phi$ is a sampling parameter. The series $N_t$ has some initial value $N_0$
and one-step conditional distributions are defined by
\begin{equation}
	N_{t+1}=R N_t\exp(-N_t+e_{t+1}),
\end{equation}
where $R$ is a growth parameter and $e_t\sim N(0,\sigma^2)$  is an independent environmental noise series.
We write $\theta = (\theta_1,\theta_2,\theta_3)^\top=(\log \phi, \log R, \log \sigma)^\top$ for the parameters. 
For our ``observed" data in this example we consider
a simulated time series of length $T=100$ with parameter
$\theta=(12,0.01,-0.75)^\top$.  We use the summary statistics used
in \cite{wood10} consisting of a combination of marginal distribution summaries, 
autocorrelation values, parameter estimates from an auxiliary autoregressive 
model and the number of zeros observed.  

We will consider three different prior density ratio classes. 
In all three cases, 
the upper bound $u(\theta)$ will be a uniform density for
$\theta$ on the range
$\left[ 11, 13 \right]\times\left[-0.02, 0.04 \right]\times\left[-2, -0.5 \right]$.
The lower bound functions $l(\theta)$ are different in the three cases.
The first case (hereafter case 1) takes $3l(\theta)=\hat{l}(\theta)$ to be a
uniform density on $[11.2,12.8]\times [-0.02,0.04] \times [-2,-0.5]$.  The second case (hereafter case 2) takes 
$3l(\theta)=\hat{l}(\theta)$ to be a uniform density 
on $[11,13]\times [-0.01,0.03]\times [-2,-0.5]$.  
The third case (hereafter case 3) takes 
$3l(\theta)=\hat{l}(\theta)$ to be a uniform density
on $[11,13]\times [-0.02,0.04]\times [-1.8,-0.7]$.
In case 1, $\hat{l}(\theta)$ and $\hat{u}(\theta)$ are densities
where $\theta_1,\theta_2$ and $\theta_3$ are independent, 
with the marginal densities for $\theta_2$ and $\theta_3$ being
the same for $\hat{l}(\theta)$ and $\hat{u}(\theta)$, whereas
the densities for $\theta_1$ are different.  So case 1 
accommodates densities where in the marginal prior
for $\theta_1$ the density can approach zero near the
edge of the support.  Case 2 is similar, but we allow
a wider range of shapes for the marginal prior density 
for $\theta_2$, whereas case 3 is more flexible in
the shape of the marginal prior density for $\theta_3$.

\begin{figure*}[htbp]
\centering
\includegraphics[width=0.9\textwidth]{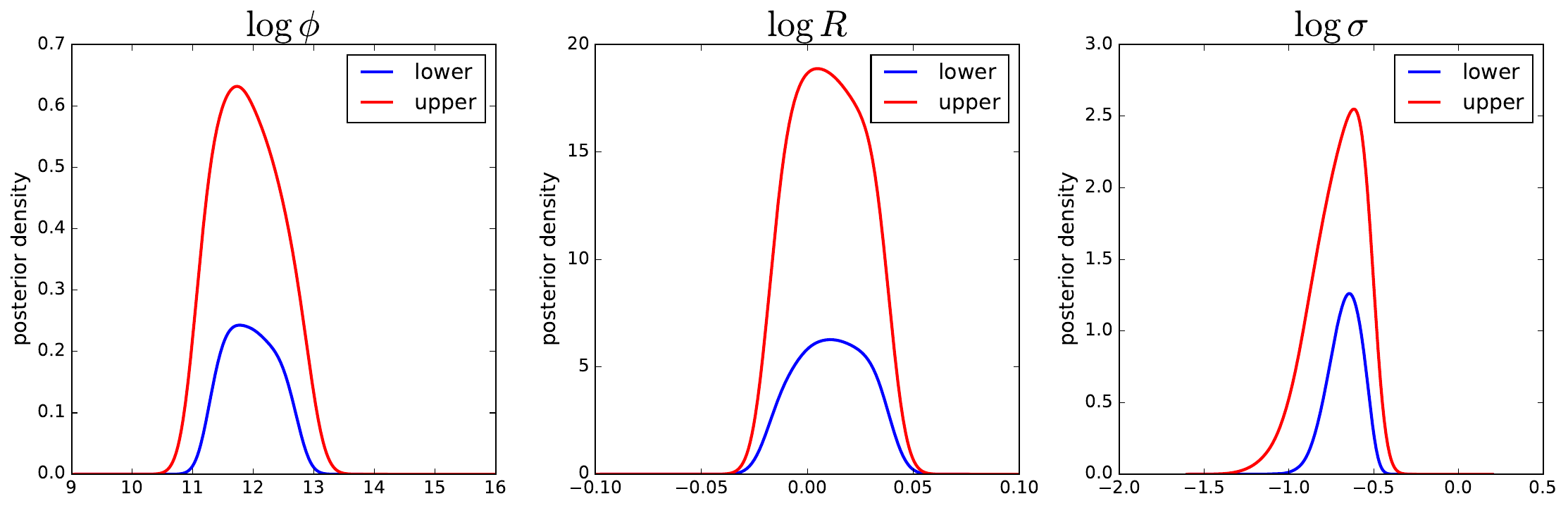}	

\vspace{1ex}
\includegraphics[width=0.9\textwidth]{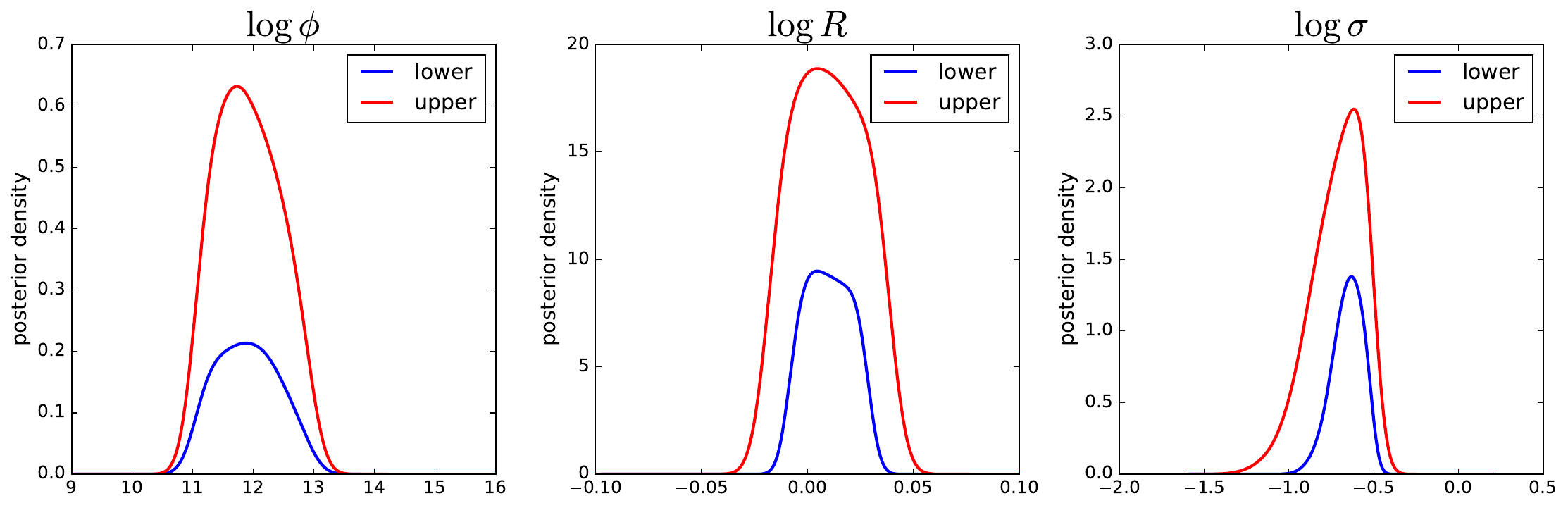}

\vspace{1ex}
\includegraphics[width=0.9\textwidth]{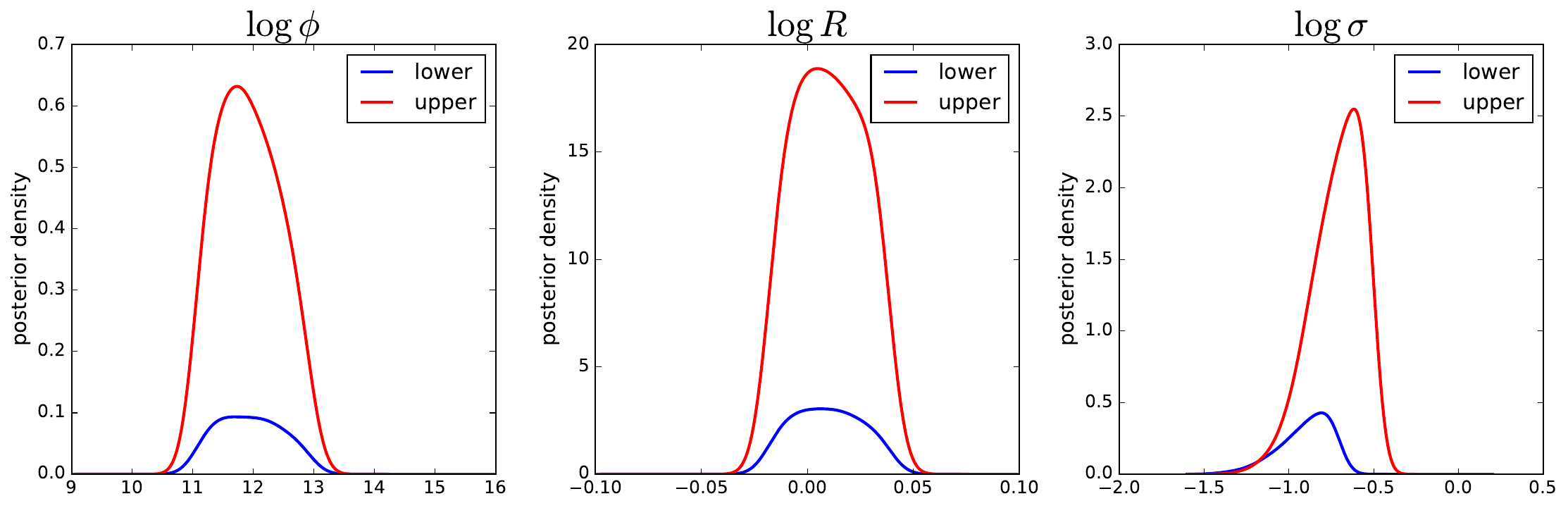}

	\caption{ Estimated upper bound and lower bound functions for
	marginal posterior density ratio classes for case 1 (first row), 
	case 2 (second row) and case 3 (third row) for the Ricker example.	}
    \label{ricker-bounds}
\end{figure*}

We use the offline training method in \cite{radev-etal-23} with $100,000$ simulated 
samples for both $\hat{l}(\theta)$ and $\hat{u}(\theta)$
to obtain posterior approximations $\widetilde{l}(\theta|S)$ and $\widetilde{u}(\theta|S)$.
Setting $S=S_{\text{obs}}$ and computing
$\widetilde{r}(S_{\text{obs}})$ using
equation (15) in the main text 
gives estimated lower and upper bounds for the posterior density
ratio class.  Figure \ref{ricker-bounds} shows estimated lower and upper bound functions $\tilde{r}(S_{\text{obs}})^{-1}\widetilde{l}(\theta_j|S_{obs})$ and $\widetilde{u}(\theta_j|S_{obs})$ for the marginal posterior density ratio classes for the three parameters, $j=1,2,3$.   
We see that the additional flexibility in the shape of the prior in the
prior density ratio class for $\log \phi$, 
$\log R$ and $\log \sigma$ for cases 1, 2 and 3 respectively is reflected in the
increased prior ambiguity in these parameters in the different cases.   
Appendix E shows diagnostics of the accuracy of the amortized posterior 
approximations.

\noindent

\section*{Appendix E:  Additional plots for the examples}

\begin{figure}[htbp]
\centerline{\includegraphics[width=2.5in]{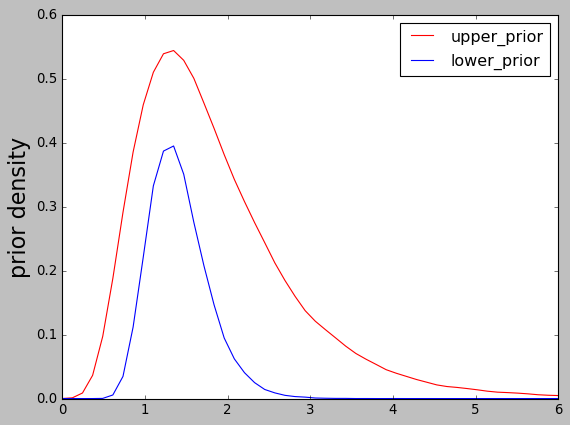}}
\caption[]{\label{toyprior} Lower and upper bound functions for prior density ratio class for the Poisson example.}
\end{figure}

For the Poisson model of Section 6.2, the lower and upper bounds are plotted in
Figure \ref{toyprior}.  

The calibration plots below provide checks for the adequacy of the amortized inference
approximations based on simulation-based calibration \citep{talts+bsvg18} for the
examples in Section 6, and the Ricker model example in Appendix D. 
The idea of simulation based calibration is to draw parameters and data from the joint
Bayesian model, then approximate posterior samples given the synthetic datasets, 
and obtain ranks of the prior samples within marginal posterior samples.  The plots below
show the difference between an empirical distribution of probability 
integral transform (PIT) values and a uniform distribution, together with simultaneous 
confidence bands.  If the line is outside the bands
this indicates inadequacy of the computational approximation \citep{sailynoja+bv22}.  

\begin{figure}[H]
    \centering
    \begin{subfigure}[t]{0.25\linewidth}
        \centering
        \includegraphics[width=\linewidth]{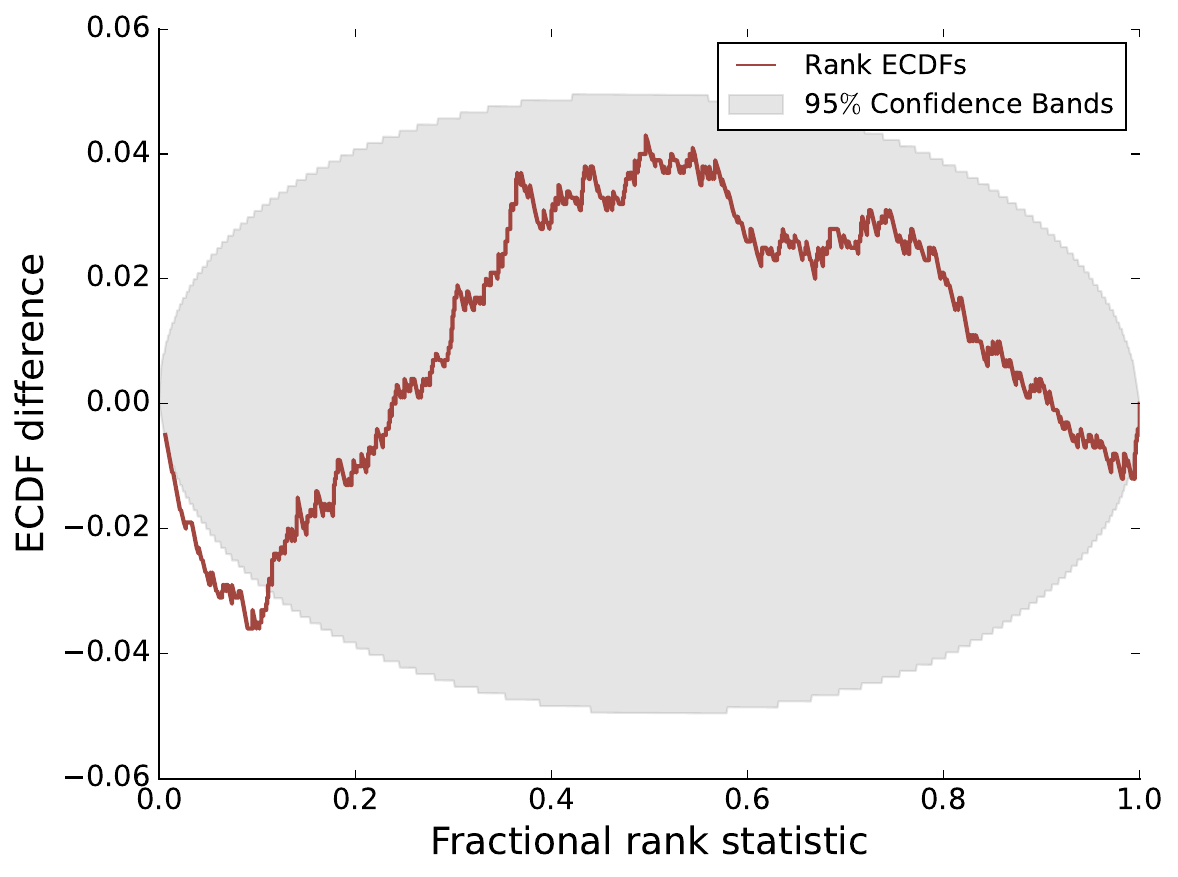}
        \par
        \includegraphics[width=\linewidth]{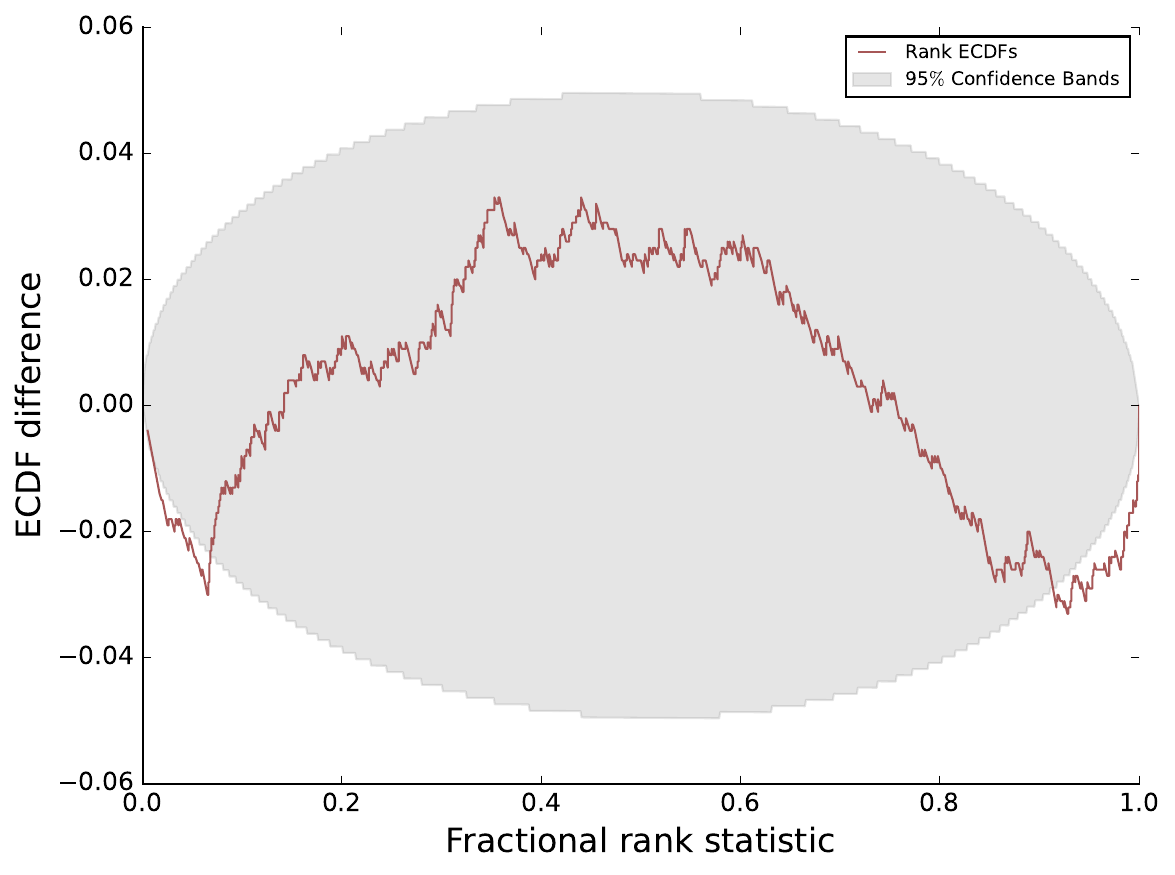}
        \caption{$\bar{y}$}
    \end{subfigure}\hfill
    \begin{subfigure}[t]{0.25\linewidth}
        \centering
        \includegraphics[width=\linewidth]{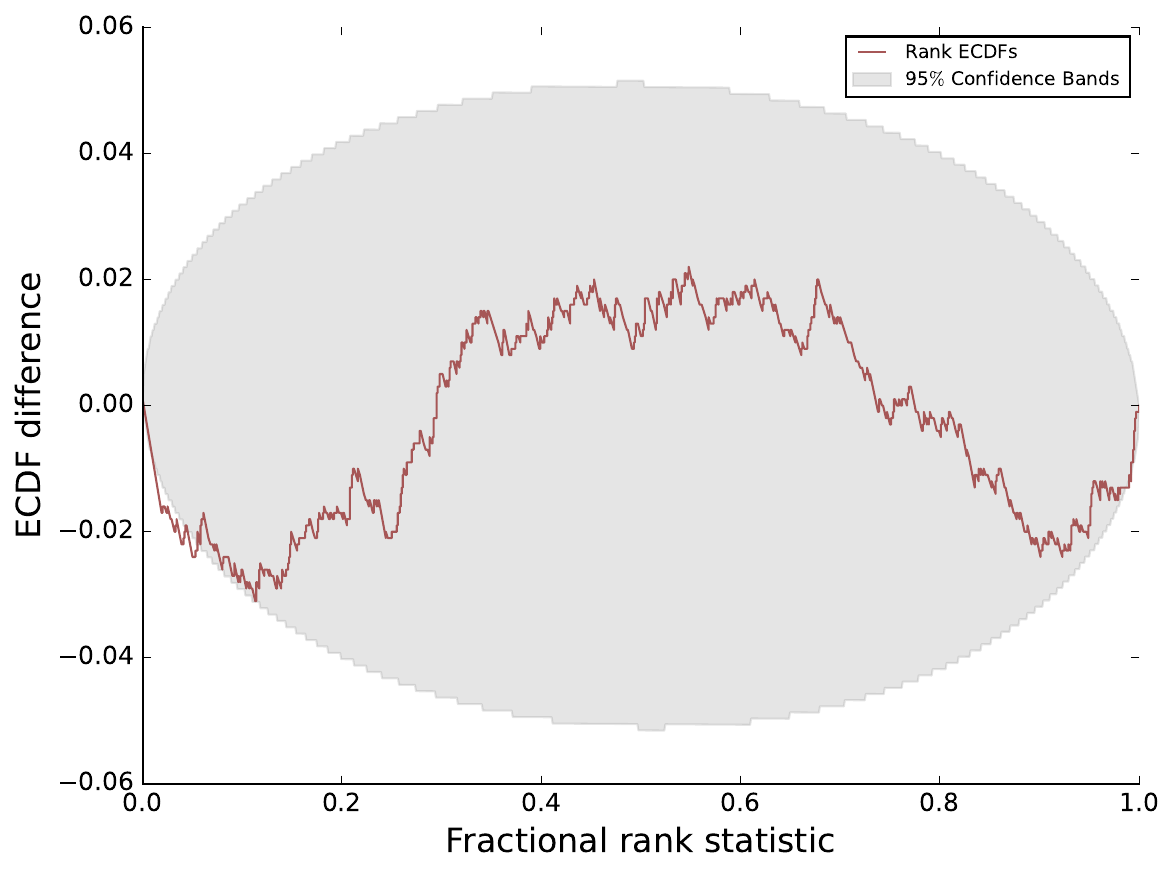}
        \par
        \includegraphics[width=\linewidth]{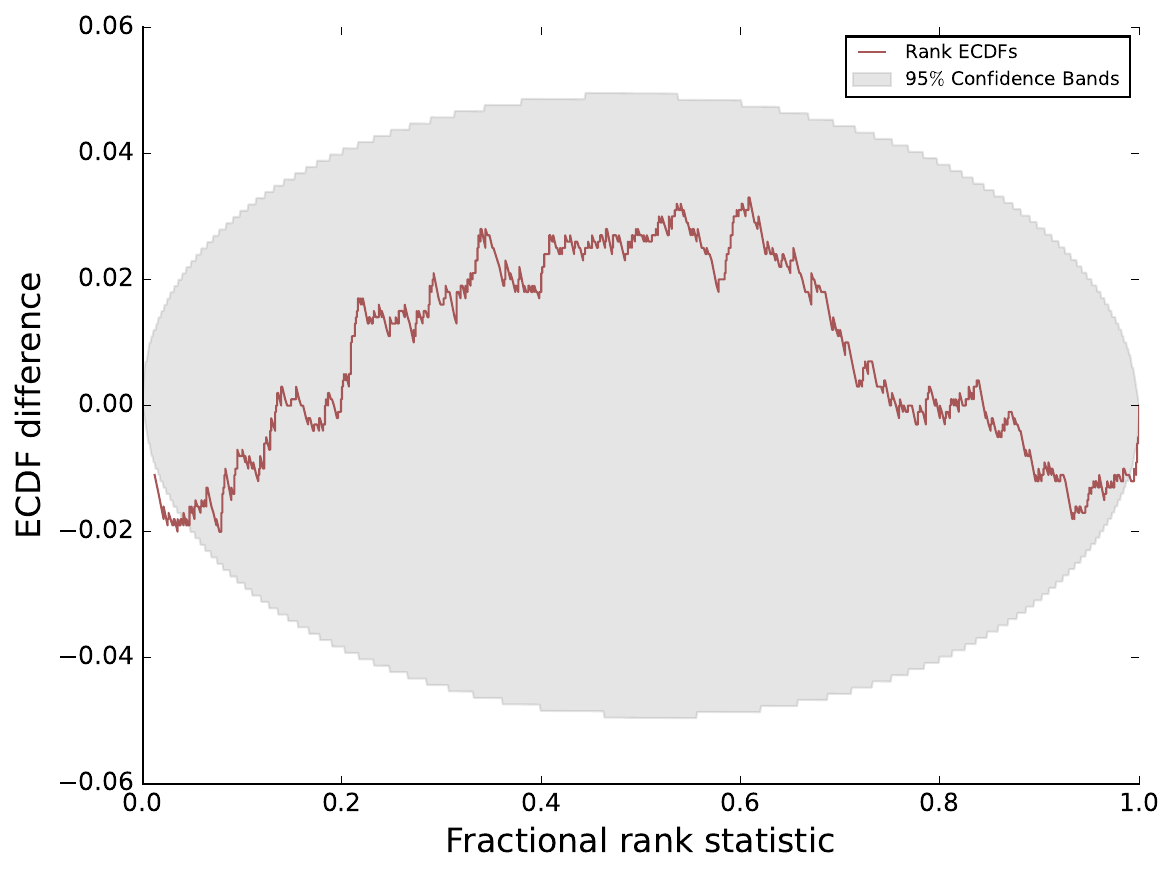}
        \caption{$(\bar{y},s^2)$}
    \end{subfigure}\hfill
    \begin{subfigure}[t]{0.25\linewidth}
        \centering
        \includegraphics[width=\linewidth]{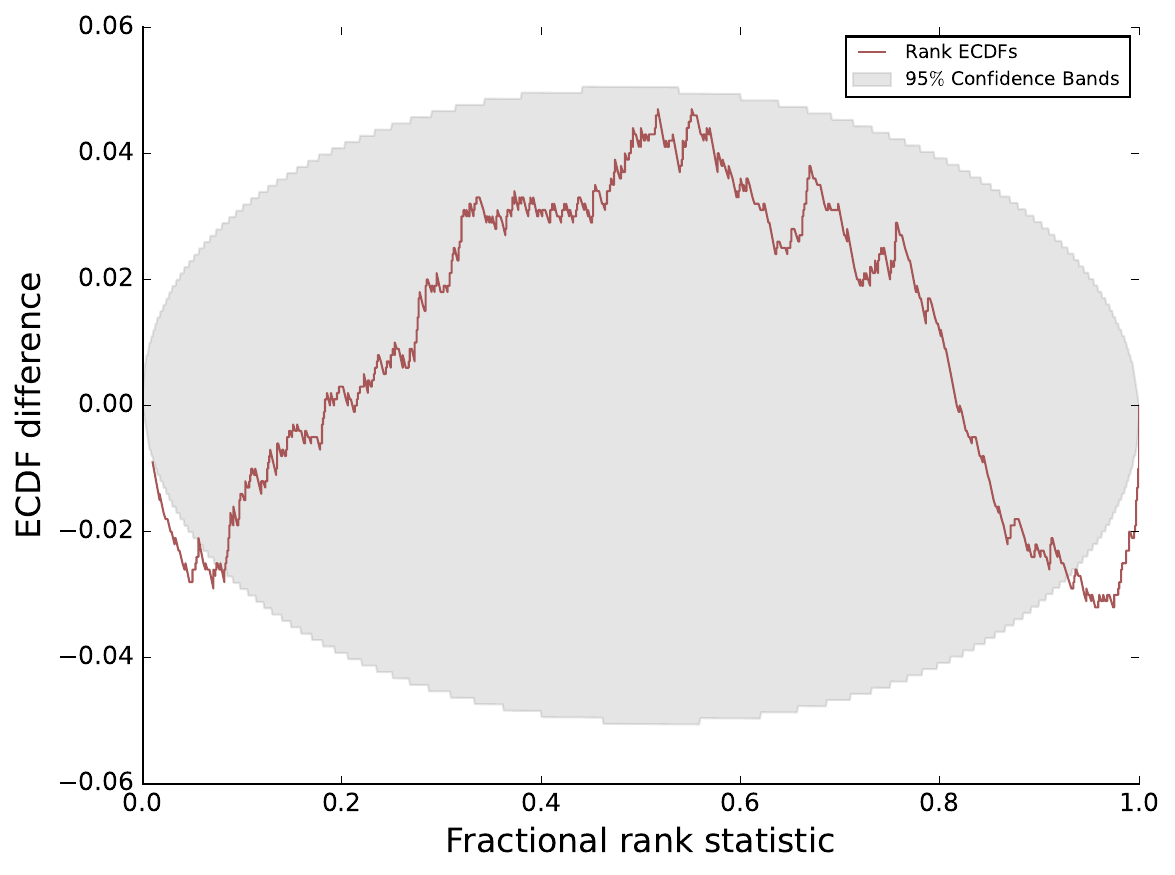}
        \par
        \includegraphics[width=\linewidth]{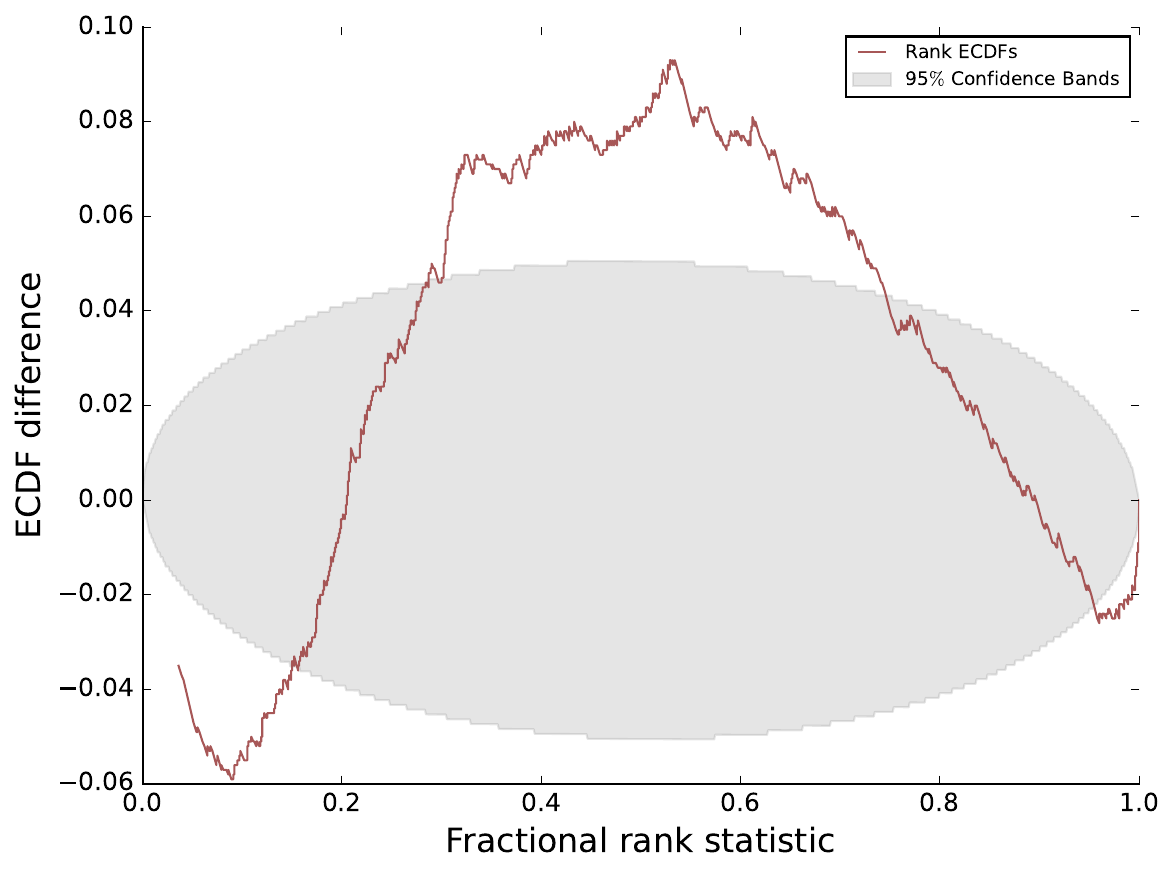}
        \caption{$s^2$}
    \end{subfigure}

    \caption{Calibration plots for diagnosing accuracy of amortized
    posterior computation for different summary statistic choices
    for the Poisson example. The first row is for the lower bound
    and the second row is for the upper bound.}
    \label{toyexample-diag}
\end{figure}

\begin{figure}[htbp]
	\centering

		\begin{minipage}[t]{0.8\linewidth}
			\centering
\includegraphics[width=4.5in]{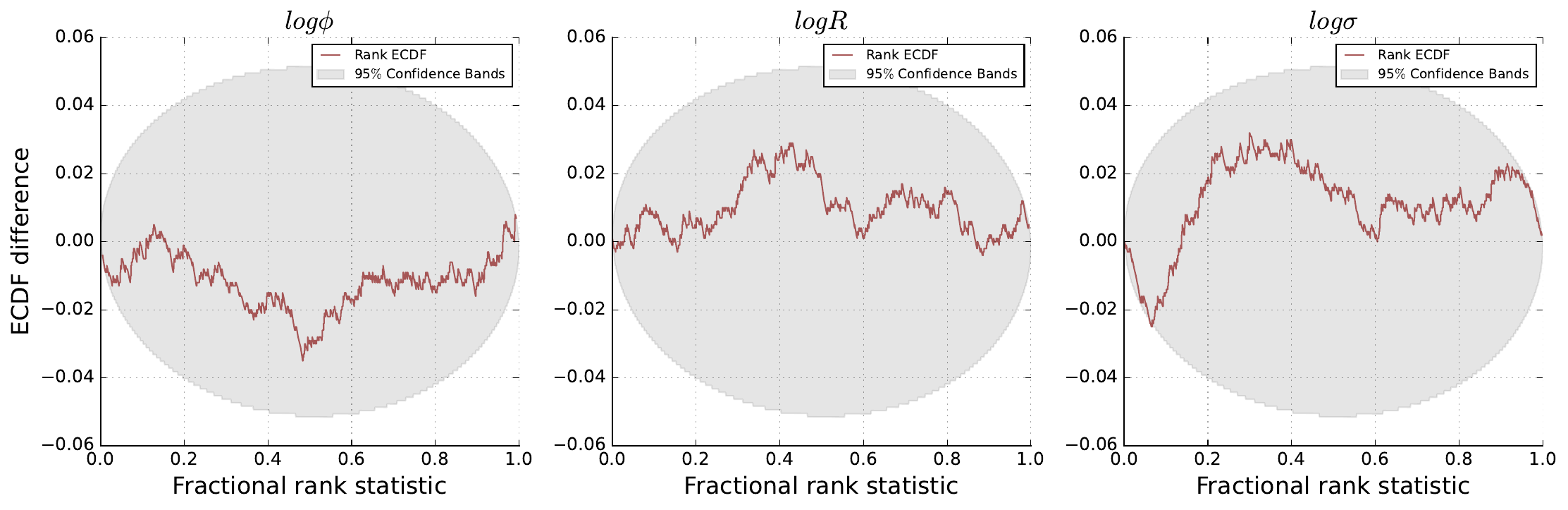}
   \\
   \includegraphics[width=4.5in]{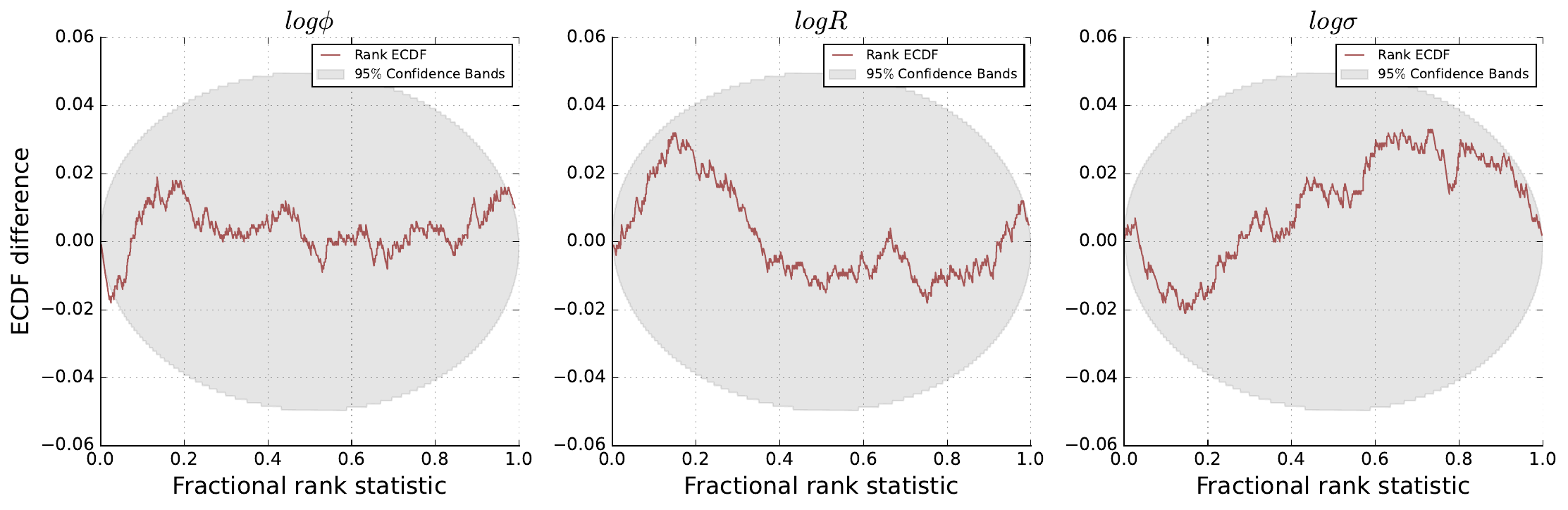}
		\end{minipage}%

	\centering
	\caption{ Calibration plots for diagnosing accuracy of amortized
	posterior computation for case 1 for lower and upper bounds for
	the Ricker example. 	The first row is for the lower bound and 
	the second row is for the upper bound.}
    \label{ricker-phi-diag}
\end{figure}

\begin{figure}[H]
	\centering

		\begin{minipage}[t]{0.8\linewidth}
			\centering
\includegraphics[width=4.5in]{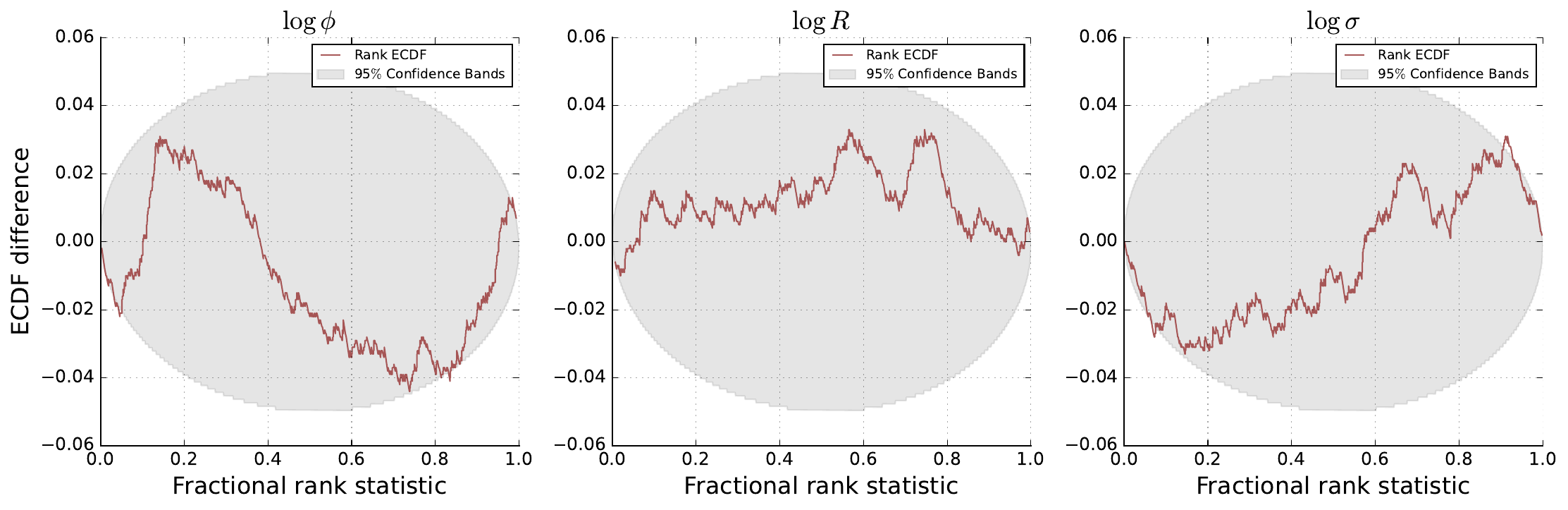}
   \\
   \includegraphics[width=4.5in]{ricker-phi-upper-diag1.pdf}
		\end{minipage}%
	
	\centering
	\caption{ 
	Calibration plots for diagnosing accuracy of amortized
	posterior computation for case 2 for lower and upper bounds for
	the Ricker example. 	The first row is for the lower bound and 
	the second row is for the upper bound.}
    \label{ricker-r-diag}
\end{figure}

\begin{figure}[htbp]
	\centering
		\begin{minipage}[t]{0.8\linewidth}
			\centering
\includegraphics[width=4.5in]{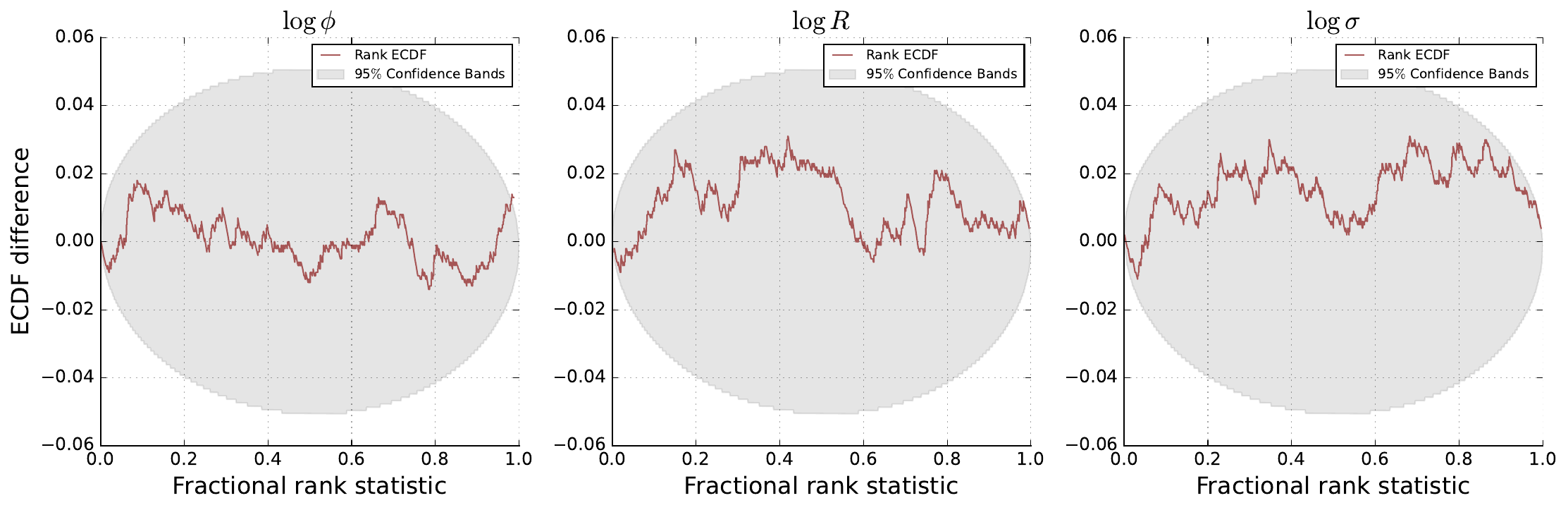}
   \\
   \includegraphics[width=4.5in]{ricker-phi-upper-diag1.pdf}
		\end{minipage}%
	
	\centering
	\caption{Calibration plots for diagnosing accuracy of amortized
	posterior computation for case 3 for lower and upper bounds for
	the Ricker example. 	The first row is for the lower bound and 
	the second row is for the upper bound.}
    \label{ricker-sigma-diag}
\end{figure}

\begin{figure}[htbp]
	\centering

		\begin{minipage}[t]{0.8\linewidth}
			\centering
\includegraphics[width=4.5in]{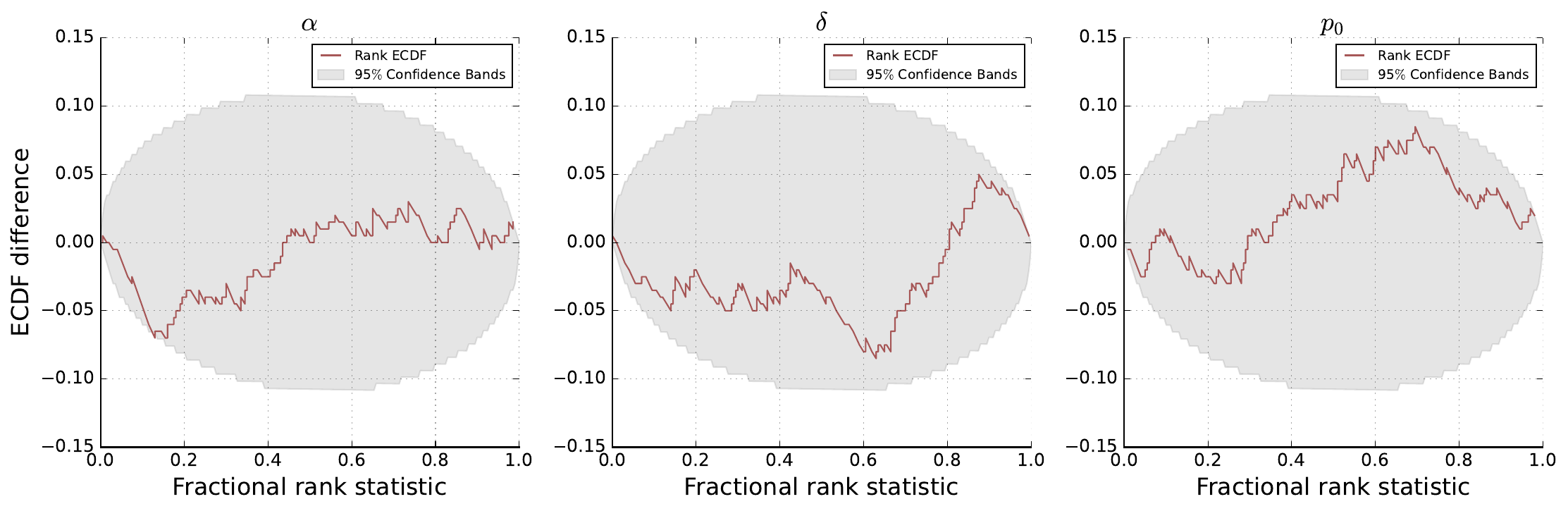}
   \\
   \includegraphics[width=4.5in]{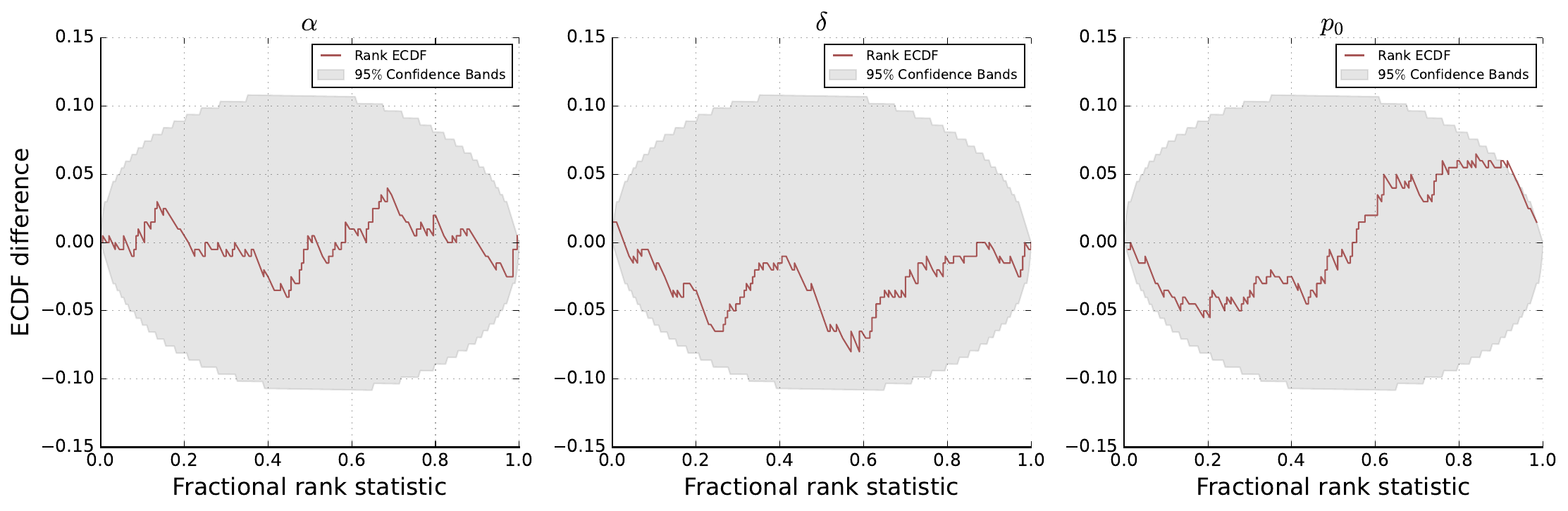}
		\end{minipage}%
	
	\centering
	\caption{Calibration plots for diagnosing accuracy of amortized
	posterior computation for case 1 for lower and upper bounds for
	the Fowler's toad example.  The first row is for the lower bound and 
	the second row is for the upper bound.}
    \label{toad-alpha-real-diag}
\end{figure}

\begin{figure}[htbp]
	\centering
		\begin{minipage}[t]{0.8\linewidth}
			\centering
\includegraphics[width=4.5in]{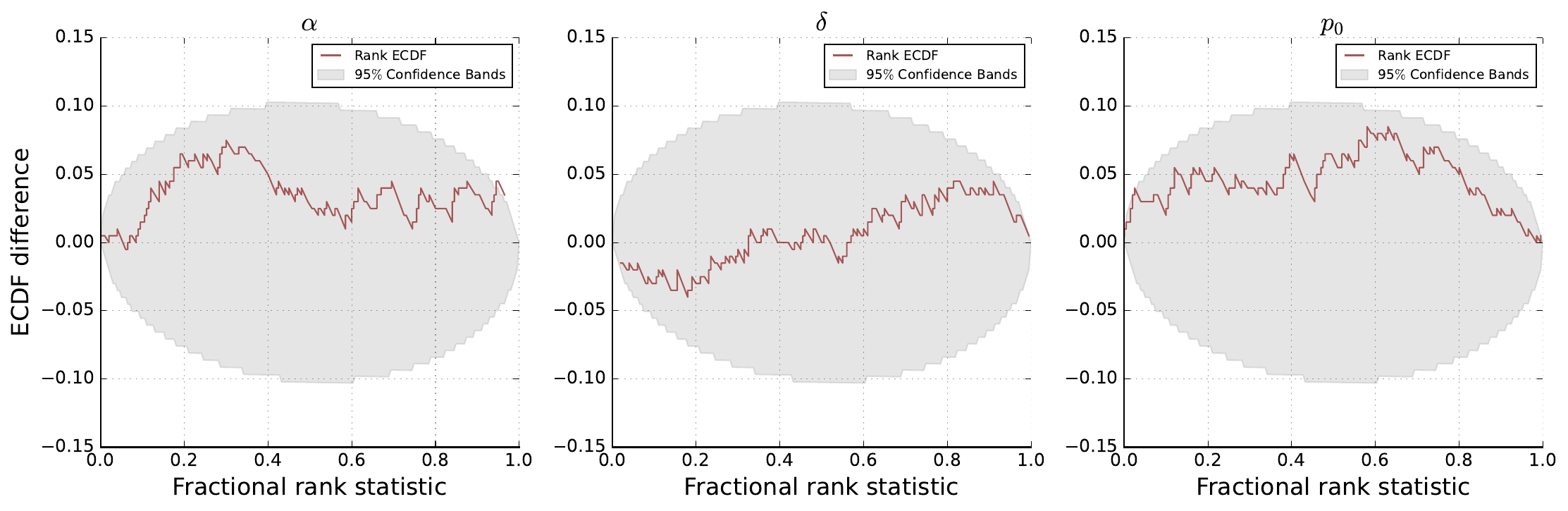}
   \\

   \includegraphics[width=4.5in]{toad-delta-upper-diag1.pdf}
   \\
		\end{minipage}%
	
	\centering
	\caption{
	Calibration plots for diagnosing accuracy of amortized
	posterior computation for case 2 for lower and upper bounds for
	the Fowler's toad example.  The first row is for the lower bound and 
	the second row is for the upper bound.}
    \label{toad-delta-real-diag}
\end{figure}

\begin{figure}[htbp]
	\centering
	
		\begin{minipage}[t]{0.8\linewidth}
			\centering
			\includegraphics[width=4.5in]{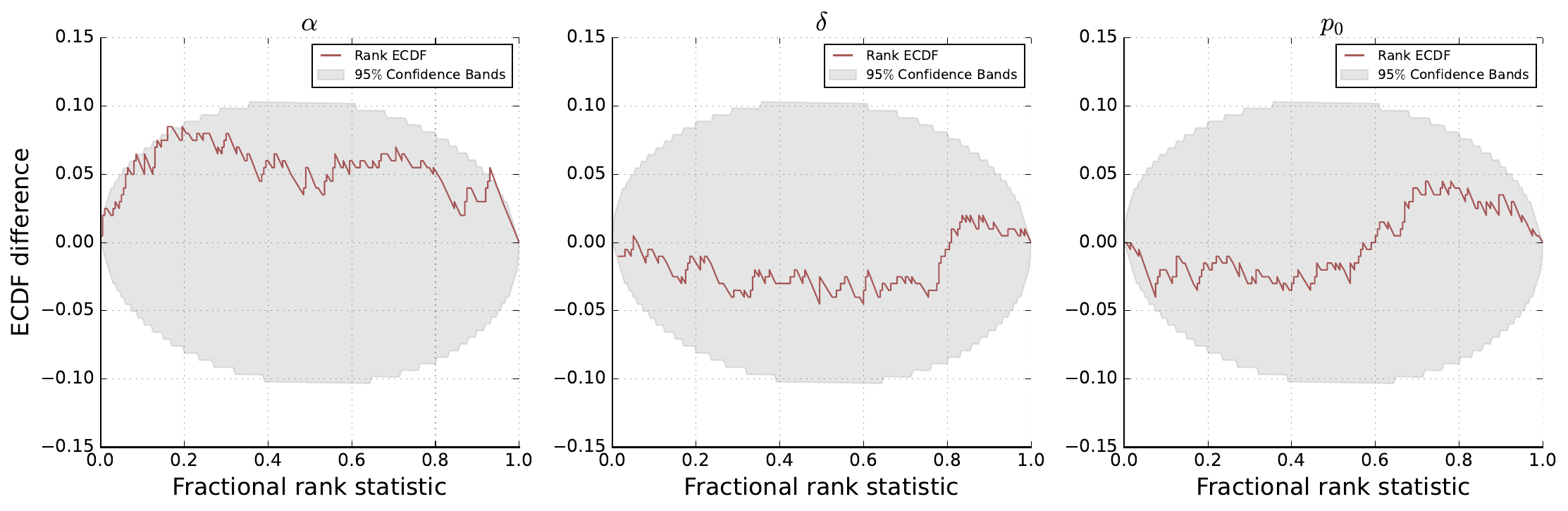}
   \\
   \includegraphics[width=4.5in]{ toad-delta-upper-diag1.pdf}
   
		\end{minipage}%
	
	\centering
	\caption{Calibration plots for diagnosing accuracy of amortized
	posterior computation for case 3 for lower and upper bounds for
	the Fowler's toad example.  The first row is for the lower bound and 
	the second row is for the upper bound.}
    \label{toad-p0-real-diag}
\end{figure}


\bibliography{density-ratio-class}

\bibliographystyle{apalike}

\end{document}